\documentclass{aa}  
\usepackage{graphicx}
\usepackage{txfonts}

\usepackage[normalem]{ulem}
\usepackage{xcolor} %
\usepackage{cprotect}

\newcommand\nube{{\em Nube}}

\begin{document}

\title{Constraining the shape of dark matter haloes using only starlight}
\subtitle{I. A new technique and its application to the galaxy {\em Nube}}

\author{Jorge S\'anchez Almeida\inst{1,2}
       \and
       Ignacio Trujillo\inst{1,2}
       \and
       Mireia Montes\inst{3}
       \and
       Angel R. Plastino\inst{4}
       }
       \institute{
       {Instituto de Astrof\'\i sica de Canarias, La Laguna, Tenerife, E-38200, Spain}\\
       \email{jos@iac.es,itc@iac.es}
       \and
       Departamento de Astrof\'\i sica, Universidad de La Laguna
       \and
       {Institute of Space Sciences (ICE, CSIC), Campus UAB, Carrer de Can Magrans, s/n, 08193 Barcelona, Spain}\\
       \email{mireia.montes.quiles@gmail.com}
       \and
       {CeBio y Departamento de Ciencias B\'asicas, 
       Universidad Nacional del Noroeste de la Prov. de Buenos Aires, 
       UNNOBA, CONICET, Roque Saenz Pe\~na 456, Junin, Argentina}\\
       \email{arplastino@unnoba.edu.ar}
       }

   \date{Received \today; accepted \dots}

   \abstract{
     We present a new technique to constrain the gravitational potential of a galaxy from the observed stellar mass surface density alone under a number of assumptions. It uses the classical Eddington Inversion Method to compute the phase-space distribution function (DF) needed for the stars to reside in a given gravitational potential. In essence, each potential defines a set of density profiles, and it is the expansion of the observed profile in this database that provides the DF. If the required DF becomes negative then the potential is inconsistent with the observed stars and can be discarded. It is particularly well-suited for analyzing low-mass low surface brightness galaxies, where photometric but not spectroscopic data can be obtained. The recently discovered low surface brightness galaxy \nube\ was used to showcase its application. For the observed \nube's stellar core to be reproduced with non-negative DF, cuspy NFW (Navarro, Frenk, and White)  potentials are highly disfavored compared with potentials having cores (Schuster-Plummer or $\rho_{230}$). The method assumes the stellar system to have spherical symmetry and isotropic velocity distribution, however, we discuss simple extensions that relax the need for isotropy  and may help to drop the spherical symmetry assumption. 
   }


   \maketitle
%

\section{Introduction}\label{sec:intro}

  In the standard cosmological model, the dark matter (DM) is made of cold collisionless particles (CDM). They evolve under their own gravity to form halos following the canonical NFW profile  (after Navarro, Frenk, and White \citeyear{1997ApJ...490..493N}), where the mass density profile increases with decreasing radius ($r$) approximately as $r^{-1}$. These predicted {\em cuspy} profiles are seldom observed \citep[e.g.,][]{2017Galax...5...17D,2017ARA&A..55..343B,2019A&ARv..27....2S} since the inferred DM haloes tend to show a constant central density or {\em core}. The formation of  cores is naturally accommodated within the standard cosmological model since purely baryonic processes move gas around leading to the transformation of the overall potential and redistributing the DM particles. In the case of dwarf galaxies, the energy turning DM cusps into cores is provided by the star-formation \citep[e.g.,][]{2010Natur.463..203G,2012MNRAS.421.3464P}, therefore, when the formed stellar mass is too small,  the baryon feedback alone cannot transform cuspy DM halos into cored halos and the DM haloes should remain NFW-like.
Even though the limiting mass characterizing these Halo Unevolved Galaxies (HUGs) is model dependent 
\citep[e.g.,][]{2016MNRAS.459.2573R,2024arXiv240902172K}, it approximately corresponds to $M_\star < 10^{6}\,{\rm M}_{\odot}$ \cite[e.g.,][]{2012ApJ...759L..42P,2014MNRAS.437..415D,2015MNRAS.454.2981C,2020ApJ...904...45H,2021MNRAS.502.4262J}. 
  Thus, if the DM haloes of galaxies with $M_\star \ll  10^{6}\,{\rm M}_{\odot}$ show cores, it would indicate the DM not being collisionless but fuzzy, self-interacting, warm, or other alternatives to CDM \citep[][]{1994PhRvL..72...17D,2000PhRvL..85.1158H,2000PhRvL..84.3760S, 2022arXiv220307354B,2024PhR..1054....1C}.

Traditionally,  the DM halo shapes are deduced from spatially-resolved kinematical measurements, which require time-consuming high spectral resolution spectroscopy in the optical, infra-red, or radio band.  Keeping in mind the need for large statistics to reach reliable  conclusions, through this approach it is nearly imposible to measure enough DM halos in the HUG regime to address the DM nature issue. However, the broad-band photometry needed to infer the stellar mass distribution in the HUG regime starts to be doable \citep[e.g.,][]{2021A&A...654A..40T,2021ApJ...922..267C,2024ApJ...967...72R,2024AJ....168...69Z} and will become routinely simple in the near future with instruments like the Rubin Observatory \citep{2019ApJ...873..111I} or the {\em Euclid} satellite \citep[e.g.,][]{2011arXiv1110.3193L}. Fortunately, one can use photometry alone to constrain the DM halo mass distribution using the classical Eddington Inversion Method \citep[EIM;][]{2006ApJ...642..752A,2010MNRAS.401.1091C,2023ApJ...954..153S}.

The EIM \citep[][]{1916MNRAS..76..572E,2008gady.book.....B,2018JCAP...09..040L,2021isd..book.....C} provides the distribution function (DF) in the phase space needed if an observed mass density profile happens to be immerse in an assumed gravitational potential. If the required DF becomes negative somewhere in the phase space,  it proves the observed density to be physically inconsistent with the assumed potential. Such inconsistency between potential and density happens for a combination particularly interesting in the context of deciphering the nature of DM, namely, when a stellar density with a core resides in a NFW potential \citep[see][]{2006ApJ...642..752A,2010MNRAS.401.1091C,2023ApJ...954..153S}. Stellar cores are common in dwarfs \citep[]{2020ApJ...892...27M,2021ApJ...922..267C,2021ApJ...921..125S,2022NatAs...6..659B,2024ApJ...967...72R}, and if this fact remains in the critical HUG range it would evidence the need to go beyond the standard cold DM model. We note that the original inconsistency was worked out for a particularly simplistic combination where the stars form a spherical system with isotropic velocities and residing in a NFW potential. However, this particular case seems to reflect a more general and profound inconsistency since the assumptions can be substantially relaxed and the inconsistency remains: it still holds for 
  (1) quasi stellar cores and quasi NFW potentials, where the central density is not exactly constant and the inner slope of the potential is not $-1$ \citep{2023ApJ...954..153S},
  (2) anisotropic orbits of the type expected in dwarfs \citep[isotropic at the center and radially biased in the outskirts, of the type Osipkov-Merritt or Cuddeford;][]{2010MNRAS.401.1091C,2023ApJ...954..153S}, 
  (3) Einasto potentials, also characteristic of CDM without the mathematical singularity at $r=0$ hampering NFW potentials \citep{2024RNAAS...8..167S}, and 
  (4) axi-symmetric systems, proving the inconsistency to go beyond the spherical symmetry assumption \citep{2024arXiv240716519S}.

  A first attempt to constrain the DM halo of real galaxies using EIM was carried out by \citet{2024arXiv240716519S}.  They  analyzed  noisy and incomplete data of around 100 low-mass ($M_\star$ between $10^6$ and $10^8\,M_\odot$) satellites of MW-like galaxies taken from  \citet{2021ApJ...922..267C}. Fits to the observed surface brightness profiles were compared  with a battery of gravitational potentials including NFW potentials as well as potentials stemming from cored mass distributions (expected in many alternatives to CDM). The method requires fitting the observed profiles with several analytic functions having variable inner and outer slopes. Between 40\,\% and 70\,\% of the galaxies are consistent with pure cores in the stellar mass distribution and thus inconsistent with NFW-like potentials. Unfortunately, the fitted galaxies are still too massive to be in the HUG regime and so to be conclusive on the nature of DM issue. Moreover, this first technique has the drawback of not providing the DF that best fit the data but just pointing out incompatibilities.

  A second alternative approach was followed by \citet{2024arXiv240716755S} that use EMI to claim deviations of the real DM from the CDM paradigm. They analyze 6 ultra-faint dwarfs in the HUG regime ($M_\star\sim 10^3$\,--\,$10^4\,M_\odot$) all showing a clear stellar core incompatible with a NFW potential and fully compatible with the cored potentials predicted by many alternatives to CDM.  Also based on the EIM, they use a new method to directly compute the DF by fitting the observed mass surface density profile as a superposition of density profiles characteristic of the assumed potential.  In essence, each potential defines a set of density profiles, and it is the expansion of the observed profile in this database that provides the DF. The new method is sketched in the Letter by \citet{2024arXiv240716755S}, but we feel compelled to describe the procedure in detail, which is the main purpose of the present paper.  The technique is of general application, provided the underlying assumptions are met,  and we have chosen the galaxy \nube , recently discovered by \citet{2024A&A...681A..15M}, to showcase the operation and difficulties of the new tool. The choice of \nube\ is not accidental. Although they have similar stellar masses, it is 7 times larger and 100 times dimmer that the  Small Magellanic Cloud. \nube\ is an outlier of most global scaling relations and does not seem to be predicted by modern CDM cosmological simulations. Thus, the characterization of the DM halo of \nube\ is particularly interesting in the context of CDM tests.

The paper is organized as follows: Sect.~\ref{sec:observation} describes the galaxy \nube\ and the stellar mass surface density data analyzed in the present work. Part of this description includes a Monte Carlo simulation (Appendix~\ref{app:error_center})  indicating how the determination of the center of the galaxy does not influence the inferred stellar mass density profile. This section also includes fitting the observed profile with several analytic forms (Sect.~\ref{sec:fit}).
Section~\ref{sec:derivation} puts forward the method to derive the DF. The general mathematical formulation (main Sect.~\ref{sec:derivation})  is specified to particular potentials in
Sect.~\ref{sec:eigen_sp} (Schuster-Plummer potential),  Sect.~\ref{sec:eigen_nfw} (NFW potential), and Appendix~\ref{sec:rhoAbc} ($abc$ potential). The interpretation of our formulation in terms of the classical statistical mechanics of gravitating systems is carried out in Appendix~\ref{app:classical}.
The actual implementation of the algorithm, which follows a Bayesian approach, is presented in Sect.~\ref{sec:the_actual_algorithm}, with a number of sanity checks presented in Appendixes~\ref{sec:hyper_param}, \ref{app:referee}, and \ref{sec:recovery}. Section~\ref{sec:potential_nube} describes the application to \nube\ showing how NFW potentials (cuspy) are highly disfavored compared with Schuster-Plummer potentials (cored). It is split into two subsections showing the constraints imposed by the analytic function fits (Sect.~\ref{sec:potential_nube_analytic}) and the actual application of the new algorithm (Sect.~\ref{sec:application}).
A short Sect.~\ref{sec:extensions} sketches simple extensions of the algorithm that relax the assumption of isotropic velocities. Finally, the conclusions are summarized in Sect.~\ref{sec:conclusions} and general guidelines to improve the method are given.

%

\begin{figure}
\centering
\includegraphics[width=\linewidth]{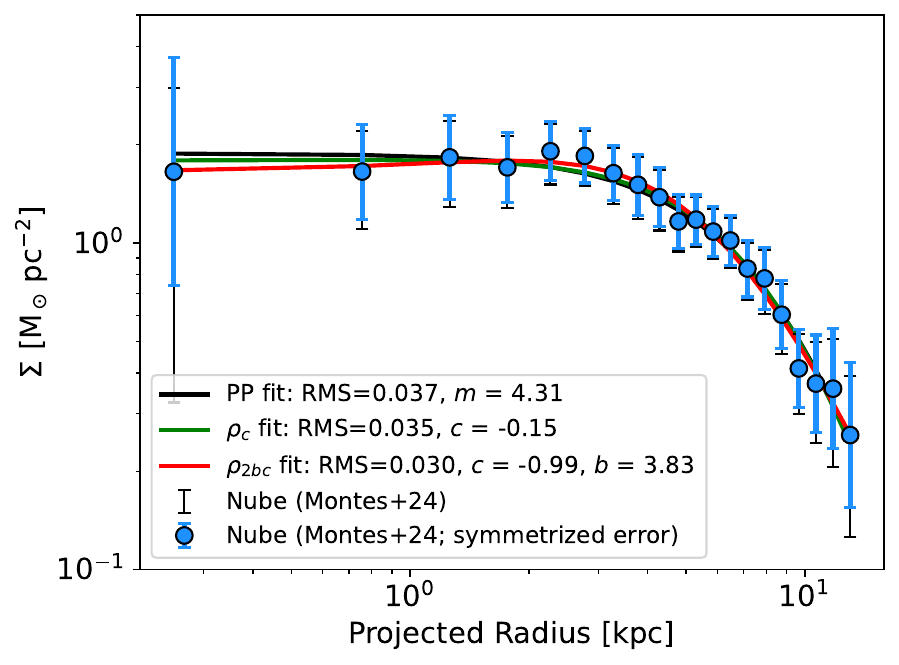}
\caption{Stellar mass surface density profile of \nube\ as observed by \citet[][the symbols with error bars]{2024A&A...681A..15M}. The figure includes three different fits to this observation: a projected polytrope (PP; the black line, with its index $m$ given in the inset) and two projected $abc$ profiles (Eq.~[\ref{eq:rhoabc}]). The inner slope ($-c$) is forced to be zero in the PP fit whereas becomes positive (i.e., $c<0$) when allowed to vary (see the inset). The gray error bars are those provided by  \citeauthor{2024A&A...681A..15M} (symmetric in a linear scale and so asymmetric in the logarithmic representation used in the figure) whereas the symmetrized ones$^{\ref{foot:1}}$ (the blue bars) are equivalent but symmetric in the logarithmic representation.}           
  \label{fig:nube}
\end{figure}
\section{Data and first analysis of {\em Nube}}\label{sec:observation}

\nube\ was discovered and characterized by \citet{2024A&A...681A..15M}.
It has a stellar mass similar to the Small Magellanic Cloud but happens to be unprecedentedly large. \nube\ is 7 times larger (effective radius of 6.9\,kpc) and 100 times fainter ($V$-band central surface brightness of 26.2\,mag\,arcsec$^{-2}$) than the typical galaxies of its mass ($M_\star\simeq 3.9\times 10^8\,M_\odot$). This makes \nube\ the most diffuse object of its class and, at present, a dwarf without any clear counterpart in the CDM cosmological simulations aimed at reproducing ultra-diffuse galaxies  \citep[][and references therein]{2024A&A...681A..15M}. Thus, the DM halo of \nube\ is particularly interesting so that \nube\ represents an excellent target to test the usefulness of the new tools.

For the sake of comprehensiveness, here we summarize the main properties of the observation and reduction. The stellar mass profile in Fig.~\ref{fig:nube} comes from a series of deep images in the Sloan $u, g, r, i$ and $z$ filters taken with HiPERCAM \citep{2018SPIE10702E..0LD,2021MNRAS.507..350D} operated at the 10-m GTC telescope (Gran Telescopio Canarias).
The apparent size of \nube\ is smaller than the field of view of the camera, allowing a reliable background subtraction. Simultaneously, it is  large enough to grant the inner core of the galaxy ($\sim$\,10\,arcsec) to be well resolved. After observing for $\sim$\,70\,min, the $g$-band image reaches a surface brightness limit $\sim$\,31\,mag\,arcsec$^{-2}$ (3$\sigma$ in areas equivalent to $10\times 10\,{\rm arcsec}^{2}$). The surface density stellar mass was inferred from the photometry in the $g$-band, with a mass-to-light ratio inferred from $g-r$ using the calibration in \citet{2015MNRAS.452.3209R}, which assumes a \citet{2003ApJ...586L.133C} initial mass function. The stellar mass surface density profile, $\Sigma(R)$, was computed from ring averages at different radial distances, up to 30\,arcsec from the center of the galaxy.

The resulting stellar mass surface density profile is shown in Fig.~\ref{fig:nube}. The error bars were calculated as a combination of Poisson noise and errors involved in subtracting the background \citep[for further details, see][]{2024A&A...681A..15M}. The individual points in the profile come from averages in rings that do not overlap, therefore, their errors are independent. Since we will fit $\Sigma(R)$ in a logarithmic scale, for convenience, the original error bars provided by \citet{2024A&A...681A..15M} are used after symmetrization in $\log\Sigma$\footnote{If $\Delta\Sigma$ is the error of $\Sigma$, then we define the {\em symmetrized} error bar in a logarithmic scale $\Delta\log(\Sigma)$ as the error obtained by error propagation, explicitly,  $ \Delta\log(\Sigma)=\log{\rm e}\,\Delta\Sigma\,\Sigma^{-1}$.\label{foot:1}}. Both the original error bars (in grey color) and the symmetric ones (in blue) are given in Fig.~\ref{fig:nube}. 

The derivation of the profile shown in Fig.~\ref{fig:nube} depends on several assumptions, whose influence on the analysis presented in the following sections was evaluated and determined to be secondary. The selection of the galaxy center was examined and found to have a negligible impact on $\Sigma(R)$ (Appendix~\ref{app:error_center}). The uncertainties in the error bar estimate was addressed in Appendix~\ref{app:referee} where we also study the effect of changing the mass-to-light ratio.

%
%
\subsection{Fitting \nube\ with simple profiles}\label{sec:fit}
The first step of the analysis was fitting $\Sigma(R)$ with simple profiles. The result is also included in Fig.~\ref{fig:nube}. Polytropes are expected to describe the mass distribution when self-gravitating systems reach either thermodynamic equilibrium or another long lasting meta-stable state \citep{1993PhLA..174..384P,2020A&A...642L..14S}, and they reproduce the mass distribution in many practical instances \citep[][]{2020A&A...642L..14S,2021ApJ...921..125S}. We use the tool described in \citet{2021ApJ...921..125S} to fit \nube\ with a projected polytrope (PP). The best PP fit (the black line in Fig.~\ref{fig:nube}) does a good job reproducing the observation. Note, however, that polytropes have cores (i.e., $d\rho/dr\to 0$ when $r\to 0$) and so they are unable to follow the mild but noticeable drop of the mean $\Sigma(R)$ towards the innermost radii of \nube\ (see Fig.~\ref{fig:nube}). To be able to reproduce this drop, we also tried {with plane of the sky projections of $abc$ profiles, defined as,
\begin{equation}
  \rho_{abc} = \frac{\rho_s}{x^c(1+x^a)^{(b-c)/a}},
  \label{eq:rhoabc}
\end{equation}
where $x=r/r_s$, and $\rho_s$ and $r_s$ are scaling constants setting the volume density and the size, respectively. These profiles are commonly used to model the density of baryons or DM \citep[e.g.,][]{1990ApJ...356..359H,2006AJ....132.2685M,2014MNRAS.441.2986D} and have the advantage of encompassing the iconic NFW profile  ($a=1, b=3, {\rm ~and~} c=1$) and the $m=5$ polytrope (a.k.a. Schuster-Plummer profile, with $a=2, b=5, {\rm ~and~} c=0$). Note that $-c$ and $-b$ give the logarithmic slope of the profile in the inner and outer radii, respectively. We fit the mass surface density of \nube\ with $c$ as a free parameters and setting $a=2-c$ and $b=5-2c$, which allows the profile to seamlessly scan from a Schuster-Plummer profile to a NFW profile when   $c$ varies from 0 to 1. The result is shown as the orange line in Fig.~\ref{fig:nube}, which has $c\simeq -0.15$ and improves the root-mean-square (RMS) of the residuals with respect to the PP fit (see the inset in Fig.~\ref{fig:nube}). We also try fits allowing both the inner and outer slopes $c$ and $b$ to vary (the red line in Fig.~\ref{fig:nube}, which assumes $a=2$, fixed to minimize the number of free parameters). The fit is even better, also yielding a positive inner slope ($c\simeq -1$). The fact that the inner slope tends to be positive is makes it difficult to reproduce $\Sigma(R)$ self-consistently within any potential, as we will discuss in detail in Sect.~\ref{sec:potential_nube}.
%

%
%
\section{Method to derive the distribution function {\it \lowercase{f}}$(\epsilon)$}\label{sec:derivation}

For a spherically symmetric system of particles with isotropic velocity distribution, the phase-space DF $f(\epsilon)$ depends only on the particle energy $\epsilon$. Then, the volume density $\rho(r)$ turns out to be \citep[e.g.,][Sect.~4.3]{2008gady.book.....B},
\begin{equation}
  \rho(r) = 4 \pi \sqrt{2} \,\int_0^{\Psi(r)} \, f(\epsilon) \sqrt{\Psi(r) - \epsilon} \, d\epsilon,
  \label{eq:leading}
\end{equation}
with $\epsilon = \Psi - \frac{1}{2} v^2$ the relative energy per unit mass of each  particle, $v$ the particle's velocity, and $\Psi(r) = \Phi_0 - \Phi(r)$ the relative potential energy.  The symbol $\Phi(r)$ stands for the gravitational potential energy  and $\Phi_0$ is the gravitational potential energy evaluated at the edge of the system. The previous equation can be rewritten as
\begin{equation}
  \rho(r) = \int_0^{\epsilon_{max}}f(\epsilon)\,\xi(\epsilon,r)\, d\epsilon,
  \label{eq:master}
\end{equation}
with 
\begin{equation}
  \xi(\epsilon,r)=4 \pi \sqrt{2}\sqrt{\epsilon_{\rm max}}\sqrt{\frac{\Psi(r)}{\Psi(0)}-\frac{\epsilon}{\epsilon_{max}}} \,\,\Pi(X-r),
  \label{eq:master_mind}
\end{equation}
where $\epsilon_{max}= \Psi(0)$, $X$ is the radius implicitly defined as $\Psi(X)/\Psi(0) = \epsilon/\epsilon_{max}$, and $\Pi(x)$ represents the step function,
\def\cacab{if $x < 0$}
\def\cacac{if $x \geq 0$}
\begin{equation}
  \Pi(x) =  
  \begin{cases}
   0 & {\rm if~} x < 0 ,\\
   1 & {\rm if~} x \geq 0.
   \end{cases}
\end{equation}
Equation~(\ref{eq:master}) admits a physically revealing interpretation.  The function $\xi(\epsilon,r)$ parameterizes a family of energy dependent volume densities characteristic of the potential $\Psi$.\footnote{We note that $\xi$, as defined in Eq.~(\ref{eq:master_mind}),  has units of velocity. It could have been redefined scaling $\xi$ with a trivial constant factor to yield proper mass volumen density units. However, we have preferred  to leave it as is for formal simplicity and because, as we explain in Sect.~\ref{sec:the_actual_algorithm}, a global scaling factor in $f$ or $\xi$ does not affect neither our technique nor the results it provides.} Then, the volume density is just the superposition of these other characteristic densities with the DF parameterizing the contribution of each energy (see Eq.~[\ref{eq:master}]).  Examples of these characteristic densities for NFW and Schuster-Plummer potentials are given in Figs.~\ref{fig:mass_navarro1a} and \ref{fig:mass_navarro1b}, with their derivation worked out in Sects.~\ref{sec:eigen_nfw} and \ref{sec:eigen_sp}, respectively. The general case of an $abc$ potential is treated in Appendix~\ref{sec:rhoAbc}.
\begin{figure}
  \centering
\includegraphics[width=0.9\linewidth]{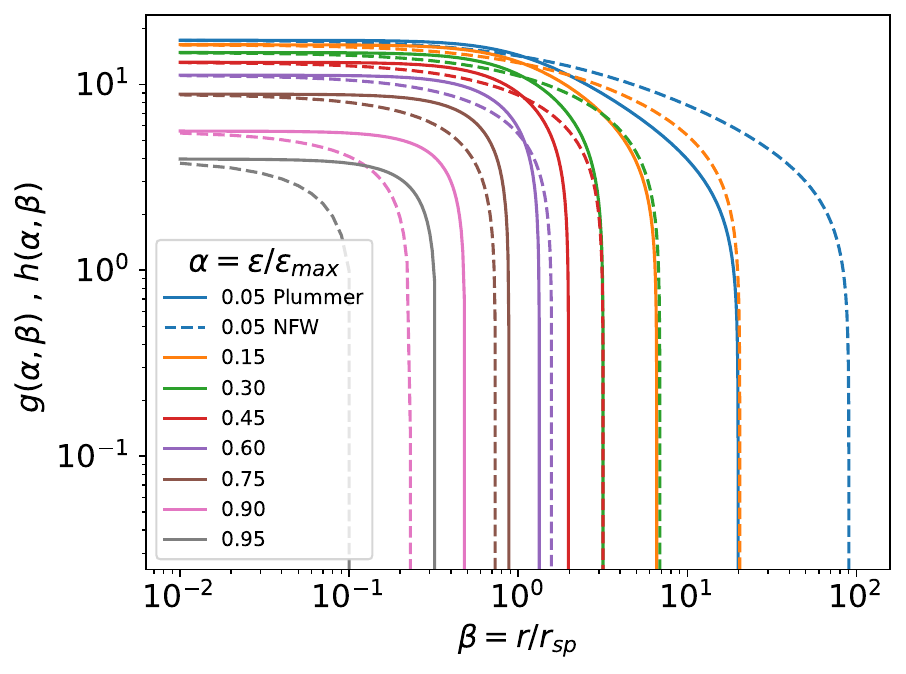} 
\caption{Characteristic density  profiles corresponding to each relative energy $\alpha =\epsilon/\epsilon_{max}$ in a Schuster-Plummer potential (the solid lines; Eq.~[\ref{eq:master_sp}]) and in a NFW potential (the dashed lines; Eq.~[\ref{eq:master_nfw}]). The symbol $\beta$ stands for the normalized radial coordinate $r/r_{sp}$. All energies contribute to the innermost regions ($\beta \ll 1$) whereas only the smallest energies contribute to the outer halo ($\beta \gg 1$).}
  \label{fig:mass_navarro1a}
\end{figure}
\begin{figure}
  \centering
\includegraphics[width=0.9\linewidth]{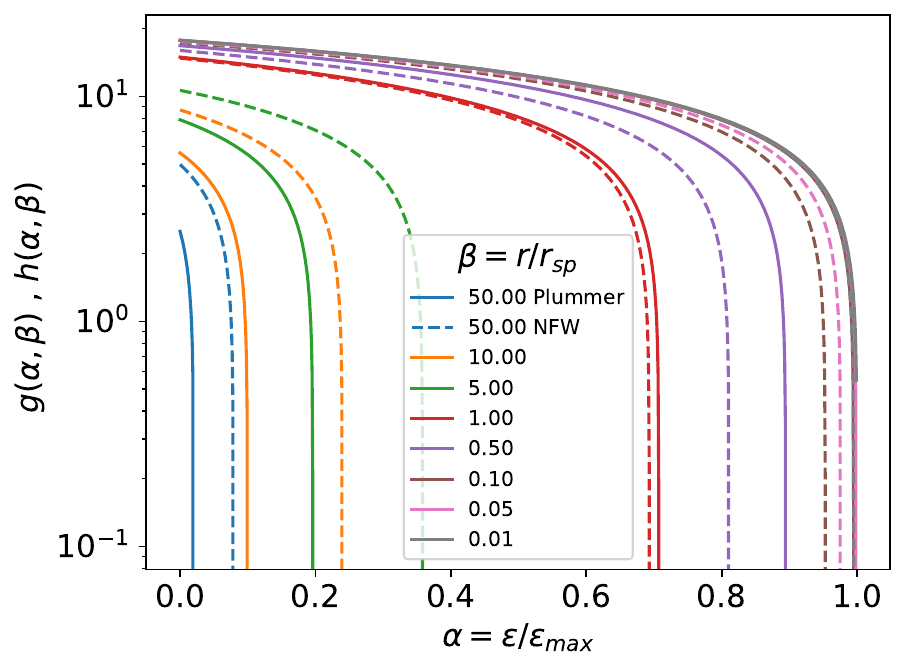} 
\caption{Similar to Fig.~\ref{fig:mass_navarro1a} but representing the contribution to the total density of the different energies ($\alpha$) at each constant radial position ($\beta$).}
\label{fig:mass_navarro1b}
\end{figure}
\begin{figure}
  \centering
\includegraphics[width=0.9\linewidth]{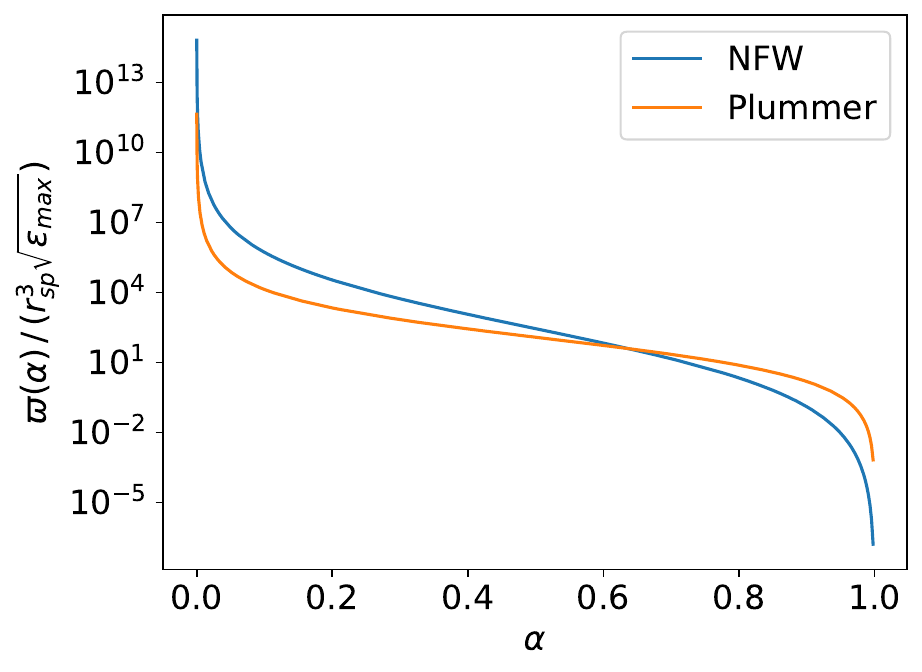} 
\includegraphics[width=0.9\linewidth]{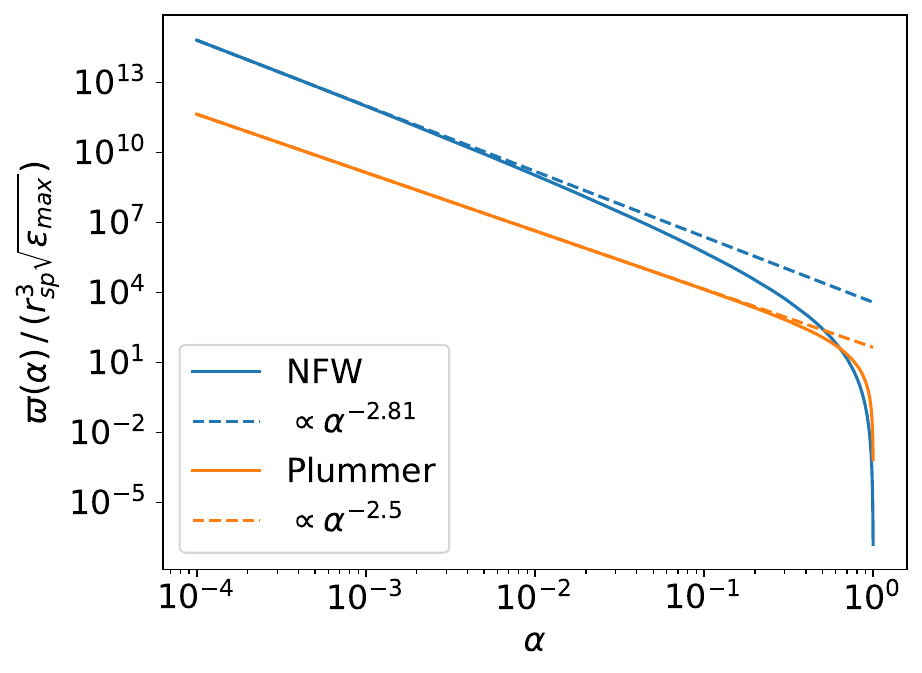} 
\caption{Specific mass corresponding to each energy (Eq.~[\ref{eq:little_mass}] with $\alpha=\epsilon/\epsilon_{max}$) for Schuster-Plummer and NFW potentials. Top panel: log-linear representation. Bottom panel: log-log representation.
Note that the specific mass diverges for low energies as a power-law of exponent -2.5, for the Schuster-Plummer potential, and around -2.81, for the NFW potential (the dashed lines labeled in the inset of the bottom panel).
}
\label{fig:mass_epsilon}
\end{figure}
Integrating Eq.~(\ref{eq:master}) over all the volume,  the total mass of the system $M$ turns out to be an integral of the masses corresponding to the different $\epsilon$, explicitly,
\begin{equation}
  M=\int_0^{\epsilon_{max}}\,f(\epsilon)\,\varpi(\epsilon)\,d\epsilon,
  \label{eq:mass_contri}
\end{equation}
with
\begin{equation}
  \varpi(\epsilon)=  4\pi\int_0^{\infty}\,\xi(\epsilon,r)\,r^2d r.
  \label{eq:little_mass}
\end{equation}
The variable $\varpi(\epsilon)$ is shown in Fig.~\ref{fig:mass_epsilon} for a Schuster-Plummer potential and a NFW potential.  As we will show in Sects.~\ref{sec:eigen_sp} and \ref{sec:eigen_nfw} (and appears in Fig.~\ref{fig:mass_epsilon}) $\varpi(\epsilon)$ diverges when $\epsilon\to 0$ thus posing some general restriction on $f(\epsilon)$ when $\epsilon\ll  \epsilon_{max}$.
We note that the quantity $\varpi(\epsilon)$ coincides with the quantity that in classical statistical mechanics is known as the {\em density of states}. For the sake of clarity, the connections of our approach with the classical interpretation are pointed out and discussed in Appendix~\ref{app:classical}.

In principle, $f(\epsilon)$ could be retrieved using Eq.~(\ref{eq:master}) by fitting $\rho(r)$ with a linear superposition of $\xi(\epsilon,r)$. In practice, however, there is no unique way to discretize Eq.~(\ref{eq:master}) for such purpose. We approach the practical problem  expanding $f(\epsilon)$ as a polynomial of order $n$,     
\begin{equation}
  f(\epsilon)\simeq \sum_{i=3}^n\,\frac{a_{i}}{\epsilon_{max}^{3/2}}\,(\epsilon/\epsilon_{max})^i,
  \label{eq:polydef}
\end{equation}
so that
\begin{equation}
  \rho(r) \simeq \sum_{i=3}^n\,a_i\,F_i(r),
\end{equation}
with
\begin{equation}
  F_i(r) =\int_0^{1}\,\alpha^i\,\frac{\xi(\alpha\,\epsilon_{max},r)}{\sqrt{\epsilon_{max}}}\,d\alpha .
  \label{eq:master2}
\end{equation}
Note that the polynomial expansion in Eq.~(\ref{eq:polydef}) lacks the three first terms (it begins at $i=3$). This is a constraint imposed by the need to have a finite total mass since for $\epsilon\to 0$, $\varpi(\epsilon)$ diverges as $\epsilon^{-\gamma}$ with  $2 < \gamma < 3$ (see Fig.~\ref{fig:mass_epsilon} and Sects.~\ref{sec:eigen_sp} and \ref{sec:eigen_nfw}).  
We also note that the normalization   in Eq.~(\ref{eq:master2})  was  chosen so that $F_i(r)$ does not depend on $\epsilon_{max}$ (see Eq.~[\ref{eq:master_mind}]). The discretization also holds for the projection of the 3D densities in the plane of the sky, i.e., 
\begin{equation}
 \Sigma(R) \simeq \sum_{i=3}^n\,a_i\, S_i(R),
  \label{eq:master3}
\end{equation}
\begin{equation}
 S_i(R) = \int_0^{1}\,\alpha^i\,\frac{\xi_\Sigma(\alpha\epsilon_{max},R)}{\sqrt{\epsilon_{max}}}\,d\alpha,
  \label{eq:master4}
\end{equation}
where $\Sigma(R)$ and $\xi_\Sigma(\epsilon,R)$ are the 2D projection (the Abel transform) of $\rho(r)$ and $\xi(\epsilon,r)$, respectively. The symbol $R$ stands for the radial coordinate in the plane of the sky projection, as used in Sect.~\ref{sec:observation} to describe the observation of \nube . 

In the next subsections, the general expressions given above are particularized for the extreme potentials used in the work, namely, a NFW potential  (Sect.~\ref{sec:eigen_sp}) and a Schuster-Plummer potential (Sect.~\ref{sec:eigen_nfw}). The general case of the potential created by an $abc$ density is worked out in Appendix~\ref{sec:rhoAbc}.

\subsection{Case of a Schuster-Plummer potential}\label{sec:eigen_sp}
The polytrope  of index 5 is usually called Schuster-Plummer profile,
\begin{equation}
  \rho_{\rm SP}(r) = \frac{\rho_{sp}}{\left[1+(r/r_{sp})^2\right]^{5/2}},
  \label{eq:defplummer}
\end{equation}
where $\rho_{sp}$ and $r_{sp}$ are the central density and the characteristic radial scale, respectively\footnote{The same symbols for the characteristic density ($\rho_{sp}$) and radial scale  ($r_{sp}$) are used irrespectively of the functional form of the mass density defining the potential. We add a subscript $p$ to avoid confusion with the characteristic density and radius defining the stellar distribution in Eq.~(\ref{eq:rhoabc}).}. The potential produced by this {\em cored} density profile is  (e.g., \citeauthor{2023ApJ...954..153S}~\citeyear{2023ApJ...954..153S}, Eq. [A14]),
\begin{equation}
  \Psi_{\rm SP}(r) = \frac{\Psi_{\rm SP}(0)}{\left[1+(r/r_{sp})^2\right]^{1/2}},
\end{equation}
with $\Psi_{\rm SP}(0)=\epsilon_{max}=4\pi G \rho_{sp}r_{sp}^2/3$.
Using these values, Eq.(\ref{eq:master_mind}) renders,
\begin{equation}
  \xi_{\rm SP}(\epsilon,r)= \sqrt{\epsilon_{max}}\, h(\epsilon/\epsilon_{max},r/r_{sp}),
\end{equation}
with
\begin{equation}
  h(\alpha,\beta) = 4 \pi \sqrt{2}\,\sqrt{\left[\left(1+\beta^2\right)^{-1/2}-\alpha\right]}\,\Pi\left(\beta_X-\beta\right),
  \label{eq:master_sp}
\end{equation}
and $\beta_X$ defined as, 
\begin{equation}
\beta_X = \sqrt{1-\alpha^2}/\alpha.
\end{equation}
Figure~\ref{fig:mass_navarro1a} shows  $h(\alpha,\beta)$ as a function of the radial coordinate $\beta$
for a number of energies $\alpha$ (the solid lines).  All energies contribute to the innermost regions ($\beta \ll 1$) whereas only the smallest energies contribute to the outer halo ($\beta \gg 1$). Figure~\ref{fig:mass_navarro1b} also shows  $h(\alpha,\beta)$ but this time as a function of the energy for a number of radii (the solid lines).

The mass corresponding to each relative energy (Eq.~[\ref{eq:little_mass}]) happens to be,
\begin{equation}
  \varpi(\alpha\epsilon_{max})= 4\pi r_{sp}^3\sqrt{\epsilon_{max}}\,\int_0^\infty h(\alpha,\beta)\,\beta^2\,d\beta.
\end{equation}
This mass diverges at low energies  since one can prove that
\begin{equation}
  \lim_{\alpha\to 0} \varpi(\alpha\epsilon_{max})\simeq  r_{sp}^3\sqrt{\epsilon_{max}}\,\frac{\sqrt{2}\,\pi^3}{\alpha^{5/2}}.
    \label{eq:limit1}
\end{equation}
The orange solid line in Fig.~\ref{fig:mass_epsilon} shows $\varpi(\epsilon)$ computed numerically from $\xi_{\rm SP}$ using the {\tt Simpson}'s rule form {\tt Scipy} \citep{2020SciPy-NMeth}. The numerical calculation was tested against the low-energy trend given by Eq.~(\ref{eq:limit1}), which is also shown in Fig.~\ref{fig:mass_epsilon} as the dashed orange line.
%

%
\subsection{Case of a NFW potential}\label{sec:eigen_nfw}
In the case of a NFW density setting the potential,
\begin{equation}
  \rho_{\rm NFW}(r)=\frac{\rho_{sp}}{(r/r_{sp})(1+r/r_{sp})^2},
\label{eq:defnfw}
\end{equation}
it turns out to be (e.g., \citeauthor{2023ApJ...954..153S}~\citeyear{2023ApJ...954..153S}, Eq.~[A7]),
\begin{equation}
  \Psi_{\rm NFW}(r)=\Psi_{\rm NFW}(0) \frac{\ln (1+r/r_{sp})}{r/r_{sp}},
\end{equation}
with $\Psi_{\rm NFW}(0)=\epsilon_{max}=4\pi G \rho_{sp} r_{sp}^2$. Using this gravitational potential, Eq.(\ref{eq:master_mind}) renders
\begin{equation}
  \xi_{\rm NFW}(\epsilon,r)= \sqrt{\epsilon_{max}}\, g(\epsilon/\epsilon_{max},r/r_{sp}),
\end{equation}
where
\begin{equation}
  g(\alpha,\beta) = 4 \pi \sqrt{2}\,\sqrt{\left[\ln(1+\beta)/\beta-\alpha\right]} \,\Pi\left(\beta_X-\beta\right),
  \label{eq:master_nfw}
\end{equation}
with $\beta_X$ implicitly defined as, 
\begin{equation}
\ln(1+\beta_X)/\beta_X=\alpha.
\end{equation}
The  characteristic density $g(\alpha,\beta)$ is shown as a function of the radial coordinate $\beta$ in Fig.~\ref{fig:mass_navarro1a} (the dashed lines) and as a function of the energy $\alpha$ in Fig.~\ref{fig:mass_navarro1b} (the dashed lines).  

The mass corresponding to each relative energy (Eq.~[\ref{eq:little_mass}]) would be
\begin{equation}
  \varpi(\alpha\epsilon_{max})= 4\pi r_{sp}^3\sqrt{\epsilon_{max}}\,\int_0^\infty g(\alpha,\beta)\,\beta^2\,d\beta,
 \label{eq:limit2}
\end{equation}
which is shown in Fig.~\ref{fig:mass_epsilon} (the blue solid line). (The numerical integration scheme is the same as for the Schuster-Plummer potential sketched in Sect.~\ref{sec:eigen_sp}.) As it happens with the Schuster-Plummer potential, $\varpi(\epsilon)$ diverges when $\epsilon\to 0$. In this case, the numerical integration gives a mass that approximately scales as $\epsilon^{-2.81}$ (the blue dashed line in Fig.~\ref{fig:mass_epsilon}).

%
%
\begin{figure}
  \centering
\includegraphics[width=0.9\linewidth]{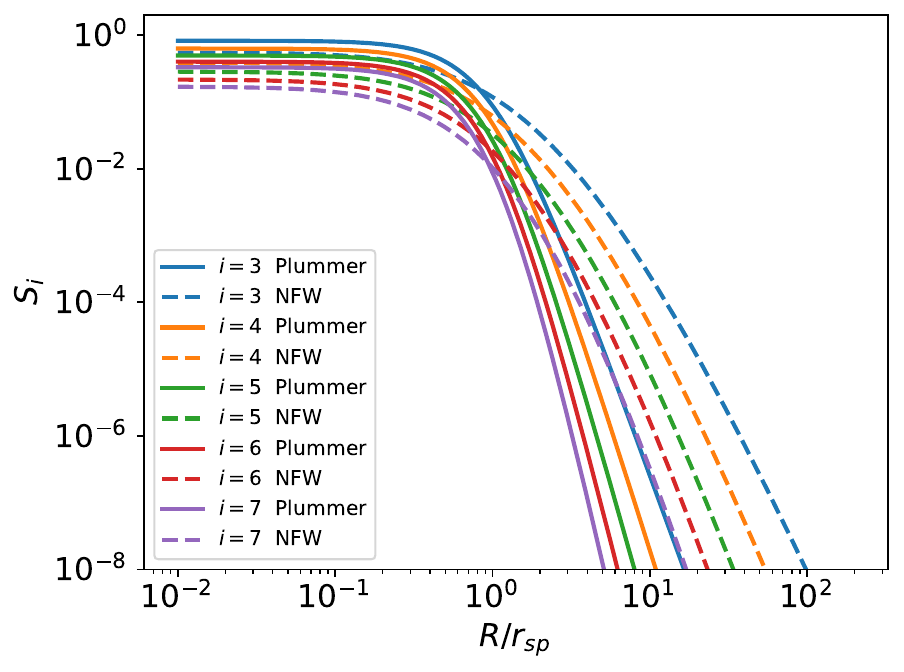} 
\caption{Characteristic functions corresponding to the polynomial expansion of the DF: see Eqs.~(\ref{eq:polydef}), (\ref{eq:master3}), and (\ref{eq:master4}). The symbol $i$ stands por the exponent of the monomial, and the solid and dashed lines correspond to the Schuster-Plummer and the NFW potentials, respectively. It begins at $i=3$ since for $i \lesssim 3$ the total mass of the resulting density profile diverges (see main text).}
\label{fig:eigendensities_poly_plot}
\end{figure}
\subsection{Algorithm to infer $f(\epsilon)$ from $\Sigma(R)$}\label{sec:the_actual_algorithm}
Except for the arbitrary scaling parameterized by $\epsilon_{max}$, Eqs.~(\ref{eq:polydef}) and (\ref{eq:master3}) provide a method to infer the DF $f(\epsilon)$ needed for a galaxy of observed mass surface density $\Sigma(R)$ to reside in a given gravitational potential. A fitting algorithm using  Eq.~(\ref{eq:master3}) provides the coefficients $a_i$ determining $f(\epsilon)$ through Eq.~(\ref{eq:polydef}). The characteristic densities in Eq.~(\ref{eq:master4}) have to be computed numerically starting from the potential in a chain requiring  at least two integrations: the Abel transform that projects the volume densities on the plane of the sky and the integral over all energies expressed by Eq.~(\ref{eq:master4}). We compute the Abel transform using the direct method implemented in the {\tt PyAbel Python} package \citep{2019RScI...90f5115H}. Then the 2nd integration is carried out using the {\tt Simpson}'s rule from {\tt Scipy} \citep{2020SciPy-NMeth}.
Several of the monomials for the Schuster-Plummer and NFW potentials are given in Fig.~\ref{fig:eigendensities_poly_plot}. All functions $S_i(R)$ show a central plateau with a power-law drop in the outskirts being more steep as the index of the monomial increases. The numerical method was tested using the  analytic solution for $S_{\frac{7}{2}}(R)$ worked out in Appendix~\ref{app:tests}.  

The free parameters retrieved from fitting $\Sigma(R)$ are the amplitudes $a_i$ together with the global radial scaling  factor setting the width of the potential $r_{sp}$, the latter making the fit non-linear. The fits were carried out using a Bayesian approach, with the log-likelihood defined as $-\chi^2/2$ where
\begin{equation}
  \chi^2 = \sum_j\left[\frac{\log\Sigma(R_j)-\log\Sigma_{m}(R_j)}{\Delta\log\Sigma(R_j)}\right]^2,
  \label{eq:chi2def}
\end{equation}
with $\Sigma(R_j)$ the observed $\Sigma(R)$ at the $j$-th radial position, $\Delta\log\Sigma(R_j)$ its error, and $\Sigma_{m}(R_j)$ the corresponding model at $R_j$. The sum includes all radii.
The posterior is explored using the   ensemble sampler for Markov Chain Monte Carlo (MCMC) {\tt emcee} \citep[][]{2013PASP..125..306F}. The best fit provided by a least squares routine that minimizes $\chi^2$ was used to initialize the exploration \citep[{\tt least\_squares} from {\tt scipy};][]{2020SciPy-NMeth}. We carried out a first unconstrained fit,  allowing $f(\epsilon)$ to vary freely. This is used as reference in all the forthcoming discussion. However, the exploration was initialized forcing the least squares routine to yield physically sensible solutions with $f(\epsilon)\geq 0$ for all $\epsilon$.   Several trial-and-error tests led us to set the final hyper-parameters used for fitting as described below. The order  of the polynomial was chosen to $n=10$, large enough to provide the flexibility needed to reproduce the inner plateau observed in \nube\ (Fig.~\ref{fig:nube}). The priors in the Bayesian analysis were chosen to be as uninformative as possible. The radial scaling $r_{sp}$ was forced to be non-negative and  $r_{sp} \le 10^3\,{\rm kpc}$ whereas the amplitudes  $a_i$ were let to vary unconstrained. Also as a prior, we asked the outermost slope of the fitted  $\log\Sigma(R)$ to be less than -2, thus preventing $\Sigma(R)$ to have infinite mass outside the observed radii. In addition, we force $f(\epsilon)\ge 0$. The posterior was explored with 32 walkers and 6000 samples.

The results reported in this paper do not depend on the exact values of the used hyper-parameters, as deduced from a number of tests detailed in Appendix~\ref{sec:hyper_param}. In these sanity checks, the  hyper-parameters are modified to  assess the effect on the interpretation of \nube . Explicitly, we tried: (1)  initializations not forced to have $f(\epsilon)\geq 0$, (2) changing the number of walkers and samples, (3) changing the order of the polynomial used for $f(\epsilon)$ ($n$ in Eq.~[\ref{eq:polydef}]), (5) constraining the amplitudes $a_i$ relative to the values of the least squares best fit, and (6) other variations referred to uncertainties in the data itself. See Appendix~\ref{sec:hyper_param} for a complete account.

\section{The gravitational potential of {\em Nube}}\label{sec:potential_nube}

\subsection{Constraints from simple fits to $\Sigma(R)$}\label{sec:potential_nube_analytic}

The observation of \nube\ in Sect.~\ref{sec:fit} show a seemingly {\em positive} inner slope,
\begin{equation}
  \omega = \lim_{R\to 0}\frac{d\log\Sigma(R)}{d\log R} > 0,
  \label{eq:innermost}
\end{equation}
which is well reproduced by profiles also having positive inner slopes in their 3D mass distribution.  In the case of  $\rho_{abc}$ profiles (Eq.~[\ref{eq:rhoabc}]) this is achieved with $c < 0$ (Fig.~\ref{fig:nube}, the red and the orange lines). The so-called  {\em cusp slope-central anisotropy theorem} by \citet{2006ApJ...642..752A} \citep[see also][]{2010MNRAS.401.1091C,2023ApJ...954..153S}  applies to spherically symmetric systems with constant  velocity anisotropy $\beta_u$ and it states that 
\begin{equation}
  c \geq 2\,\beta_u,
  \label{eq:ann}
\end{equation}
with 
\begin{equation}
\beta_u  = 1 \, - \frac{\sigma_t^2 }{2\sigma_r^2},
\label{eq:ani-param}
\end{equation}
where $\sigma_r$ and $\sigma_t$ are the radial and tangential velocity dispersions, respectively.
The form of the EIM adopted in this work (Eq.~[\ref{eq:leading}]) assumes $\beta_u=0$ so that no potential is able to reproduce the required inner positive slope (i.e., $c <0$ is inconsistent with $\beta_u=0$, according to Eq.~[\ref{eq:ann}]). This is a issue that permeates the analysis presented in subsequent sections. Obviously, the observational error bars are so large that they allow for $c=0$ or even slightly positive slope (see Fig.~\ref{fig:nube}, the green line),
but the fact that the best least squares fit is unphysical marks the study. Expanding on this argument, the light profile of \nube\ in different colors do not show the drop \citep[Fig.~4 in][and also Fig.~\ref{fig:df41_run_plot_51} below]{2024A&A...681A..15M}, which seems to appear when transforming the observed photometry into stellar mass. The impact of the uncertainty in the used mass-to-light ratio is analyzed in Appendix~\ref{app:referee}.

%
\subsection{Constraints from the full DFs }\label{sec:application}
\begin{figure}
\centering
\includegraphics[width=1\linewidth]{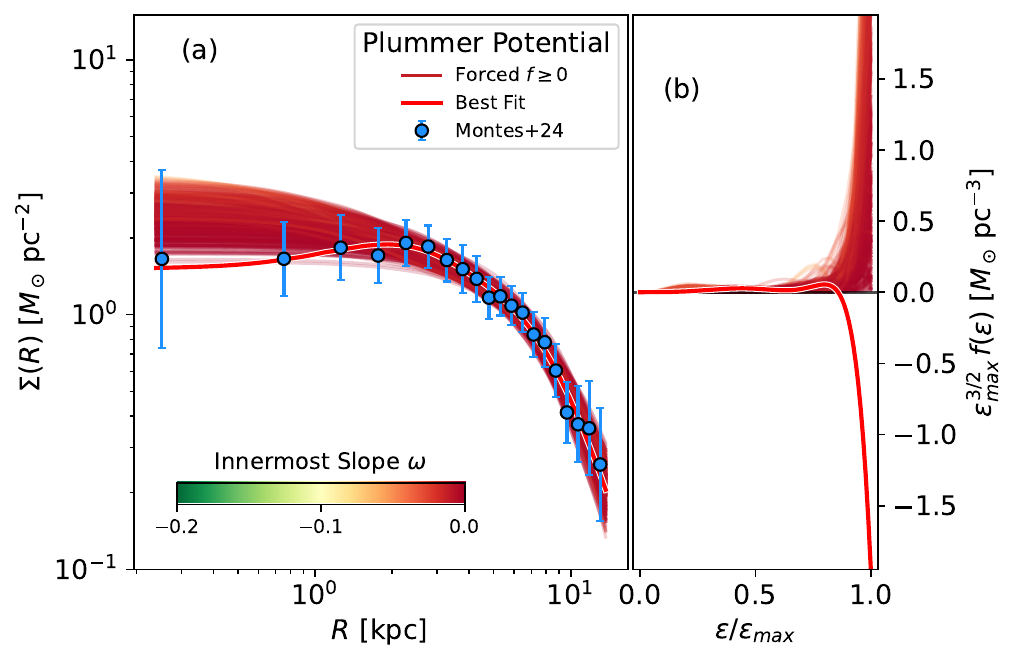} 
\caption{(a) Fit to the stellar mass surface density observed in \nube\ (the blue symbols) assumed to reside in a Schuster-Plummer potential. The red thick solid line represents the least-squares best fit with an unconstrained DF $f(\epsilon)$,   whereas the other thin lines are the fits derived from the MCMC exploration of the posterior forcing $f(\epsilon)$ to be $\geq 0$ $\forall\epsilon$ and beginning the exploration from the $f(\epsilon)\geq 0$ least squares solution. These other fits are color coded according to the innermost slope of the surface density profile ($\omega$ in Eq.~[\ref{eq:innermost}]), which always happens to be negative but close to zero.
(b) DFs required by the best fit and by the fits forced to have  $f(\epsilon)\geq 0$.  Note how the best fit requires $f(\epsilon) < 0$ and so is unphysical.  The color code is the same as in (a). 
}
\label{fig:df4_run_plot_2}
\end{figure}
Using the procedure described in Sect.~\ref{sec:derivation} and detailed in  Sect.~\ref{sec:the_actual_algorithm}, we fit the $\Sigma(R)$ of \nube\ assuming two extreme gravitational potentials; one with a core (a Schuster-Plummer potential) and another with a cusp (a NFW potential).
In addition, we also fit \nube\ assuming the gravitational potential to be generated by a $\rho_{230}$ profile (Eq.~[\ref{eq:rhoabc}] with $a= 2$, $b=3$ and $c=0$), which has a core but with the outskirts approaching a NFW profile (see Fig.~\ref{fig:df4_run_plotd}).

The result for the cored Schuster-Plummer potential are included in Fig.~\ref{fig:df4_run_plot_2}. The red solid line represents the least-squares best fit allowing for any $f(\epsilon)$ whereas the other thin lines are the fits derived from the MCMC exploration of the posterior, forcing $f(\epsilon)\geq 0$ to render physically sensible solutions and initialized with $f(\epsilon)\geq 0$ least squares solution.  We note that the best fit to the observed $\Sigma(R)$ is very good but requires $f(\epsilon) < 0$ when $\epsilon\to\epsilon_{max}$. This fact reflects the issue discussed in Sect.~\ref{sec:potential_nube_analytic} that the assumed isotropic velocity distribution is not consistent with positive inner slopes (Eq.~[\ref{eq:innermost}]).  The same exercise  with a NFW potential is included in Fig.~\ref{fig:df4_run_plot_1}. The best fit is as good as the one for the  Schuster-Plummer potential. The arrows in Fig.~\ref{fig:df4_run_plotf}c show the value of the merit function $\chi^2$  (Eq.~[\ref{eq:chi2def}]) of the best fits, and both are very similar (cf. the blue and orange arrows). However, the physically meaningful fits forcing $f(\epsilon)\geq 0$ are significantly worst in the case of a NFW potential; compare Figs.~\ref{fig:df4_run_plot_1}a and \ref{fig:df4_run_plot_2}a, the orange and blue points in Fig.~\ref{fig:df4_run_plotf}a, and the orange and blue histograms in Fig.~\ref{fig:df4_run_plotf}c.  Moreover, the innermost slope $\omega$ becomes too negative in the case of the NFW profile to be compatible with zero (Fig.~\ref{fig:df4_run_plotf}b). For the sake of completeness, Fig.~\ref{fig:df4_run_plot_3} also includes the fits resulting from assuming a $\rho_{230}$ potential. The corresponding histograms of $\omega$ and $\chi^2$ are in Fig.~\ref{fig:df4_run_plotf}, the green points and histograms.  These fits are similar and of similar quality as the fits provided by the Schuster-Plummer potential (cf. the orange with the green points and histograms in Fig.~\ref{fig:df4_run_plotf}). 
\begin{figure}
\centering
\includegraphics[width=1.\linewidth]{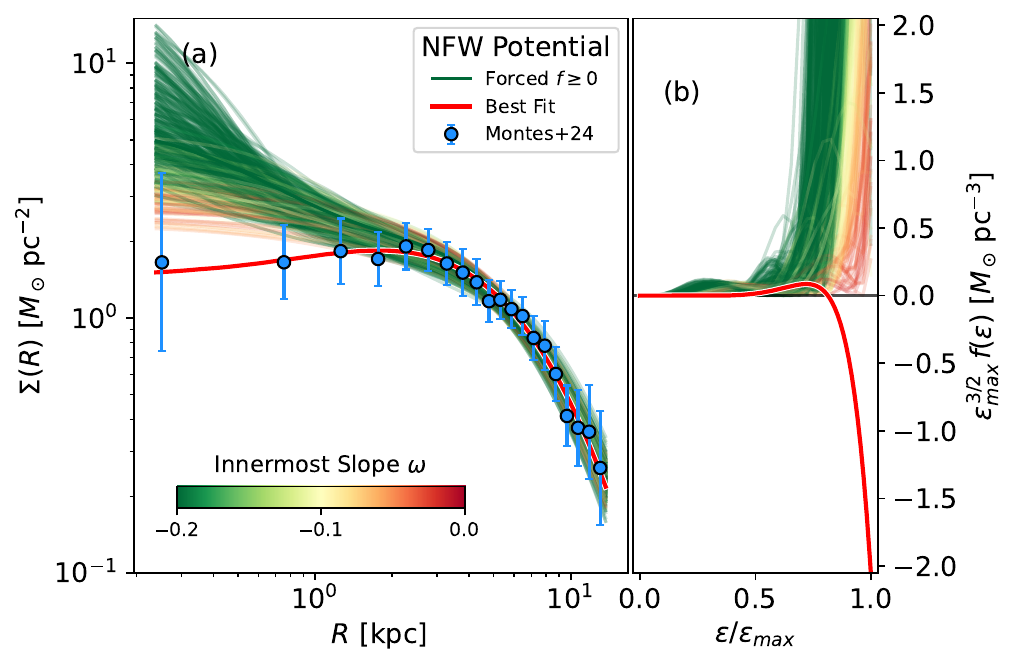} 
\caption{Similar to Fig.~\ref{fig:df4_run_plot_2} except that \nube\ is assumed to reside in a NFW potential. The best fitting function (the red thick solid line) is similar to the best fit assuming a Schuster-Plummer potential (Fig.~\ref{fig:df4_run_plot_2}a), but the physically realizable $f(\epsilon)\geq 0$ fits are clearly worst and have inner slopes differing from zero significantly (see the color code of the thin lines). 
}
\label{fig:df4_run_plot_1}
\end{figure}
\begin{figure}
\centering
\includegraphics[width=0.9\linewidth]{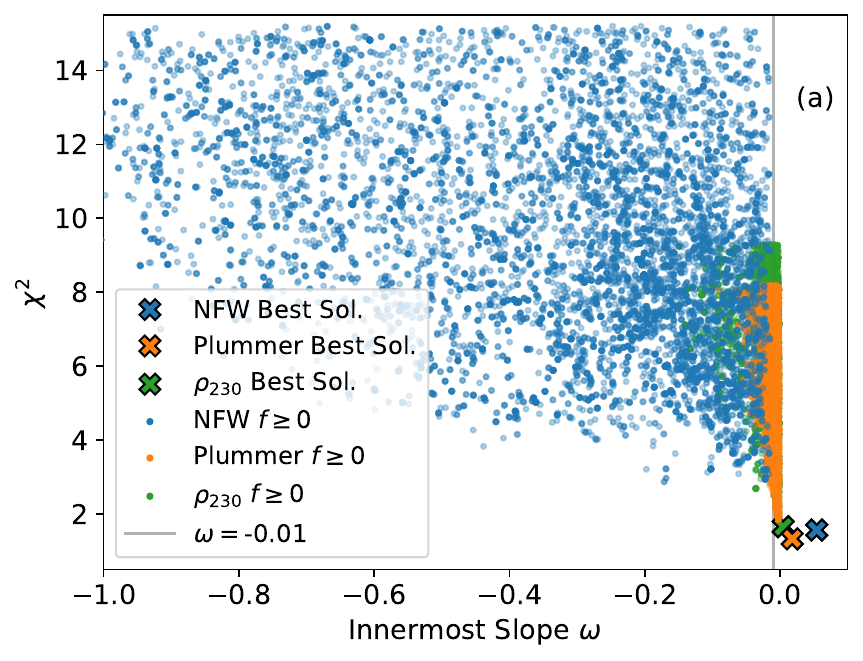}\flushleft
\includegraphics[width=1.\linewidth]{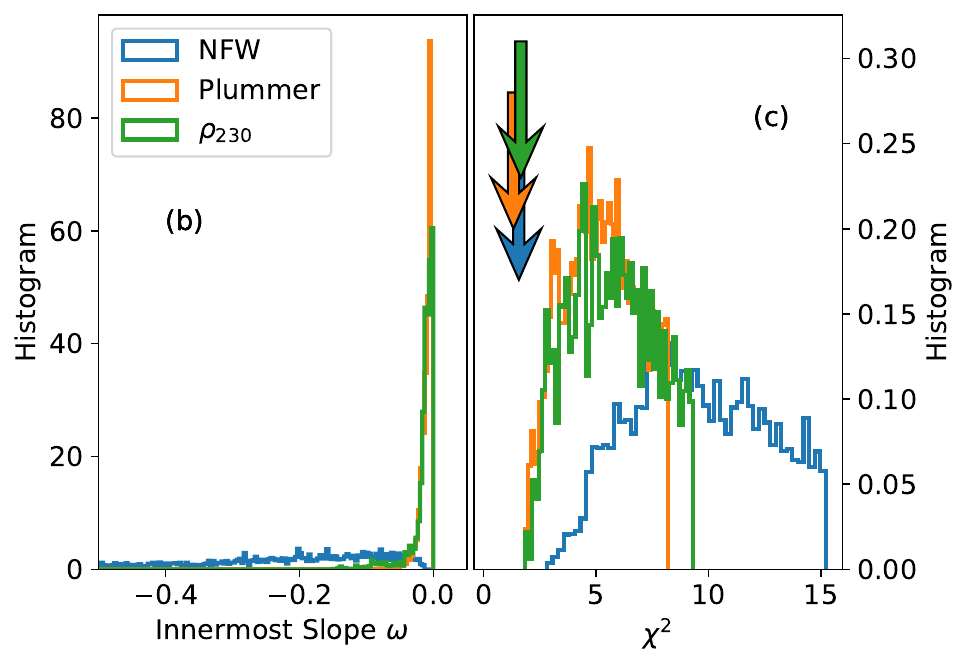} 
\caption{ (a)~Scatter plot $\chi^2$ versus  innermost slope $\omega$ of the fits for the three potentials resulting from the MCMC navigation of the posterior. The large times symbols represent the best fit obtained with unconstrained $f(\epsilon)$, which yield unphysical $f(\epsilon)< 0$. The vertical gray line marks  $\omega=-0.01$.
  (b)~Histograms of the distribution of innermost slopes for the points in (a).
  (c)~Histograms of $\chi^2$ for the points in (a). The arrows represent the $\chi^2$ of the best fit obtained with unconstrained $f(\epsilon)$. Panels (a), (b), and (c) share the same color code as indicated in the insets. 
}
\label{fig:df4_run_plotf}
\end{figure}
\begin{figure}
\centering
\includegraphics[width=1.\linewidth]{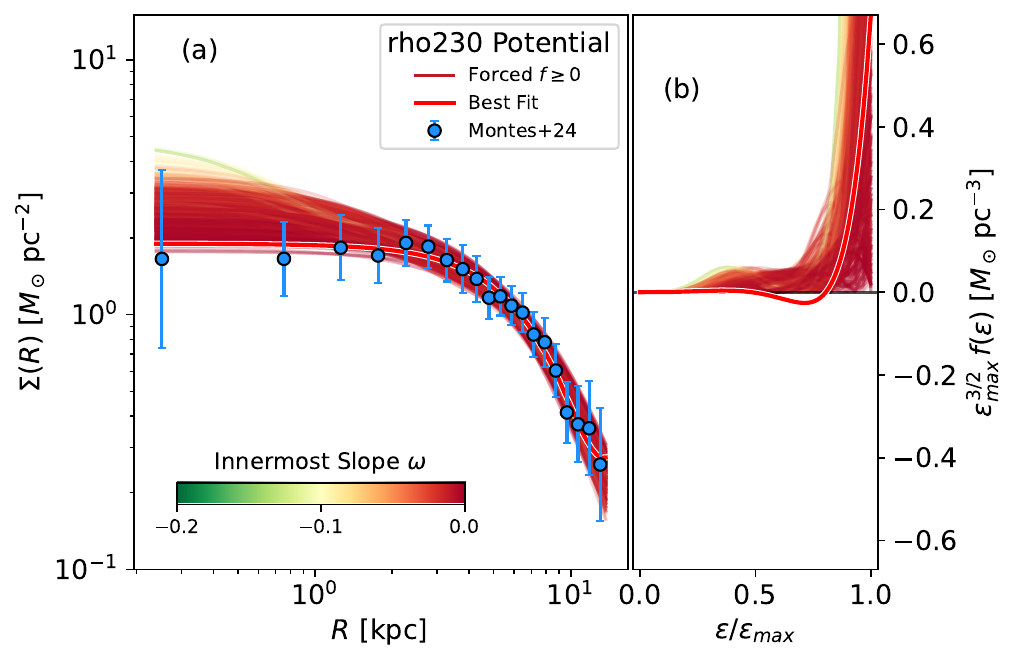} 
\caption{Similar to Fig.~\ref{fig:df4_run_plot_2} except that \nube\ is assumed to reside in a $\rho_{230}$ potential, which has a core and approaches a NFW profile in the outskirts. The best fitting function (the red thick solid line) is similar to the best fit in a NFW potential (Fig.~\ref{fig:df4_run_plot_1}),  but the physically sensible $f(\epsilon)\geq 0$ are clearly better and have a inner slope closer to zero. 
}
\label{fig:df4_run_plot_3}
\end{figure}

A way of quantifying the significant difference between the physically sensible fits based on core (Schuster-Plummer and $\rho_{230}$) and cusp (NFW) potentials comes from the histograms in Fig.~\ref{fig:df4_run_plotf}. In the case of the NFW potential, only 0.5\,\% of the points exploring the posterior have inner slope reasonably close to zero ($> -0.01$; Fig.~\ref{fig:df4_run_plotf}a). This fraction increases to 76\,\%\ and 66\,\%\ for the  Schuster-Plummer and $\rho_{230}$ potentials, respectively. Similarly, the mean value of the $\chi^2$ distribution is around 5 for Schuster-Plummer and $\rho_{230}$ and more than 8 for the NFW potential based fits. Finally, the minimum $\chi^2$ is 1.9 for both the Schuster-Plummer and the $\rho_{230}$ potentials whereas it is 1.5 times larger for the NFW potential. If this  minimum $\chi^2$ for NFW is assumed to follow a $\chi^2$ probability distribution function with 11 degrees of freedom\footnote{Which corresponds to the 20 points defining the observed \nube\ profile (Fig.~\ref{fig:nube}) minus the 9 free parameters used for fitting (Sect.~\ref{sec:the_actual_algorithm}).}, having a 1.5 times larger $\chi^2$ than the Schuster-Plummer value has a small probability of 11\,\%. In other words, the probability that the best  Schuster-Plummer $f(\epsilon)\geq 0$ fit is better than the best NFW $f(\epsilon)\geq 0$ fit is  around 89\,\%. All these statistical tests combined indicate that \nube\ is much more likely to reside in a gravitational potential with a core than with a cusp.

A number of sanity checks support that the method presented in Sect.~\ref{sec:the_actual_algorithm} works as expected when applied to known profile-potential pairs, thus supporting the above conclusions. They are discussed in Appendix~\ref{sec:sanity}. The impact of the assumed hyper-parameters of the bayesian fit, including the priors, are analyzed in Appendix~\ref{sec:hyper_param}. The effect of the uncertainties in the error assigned to \nube , which enter into the definition of the merit function (Eq.~[\ref{eq:chi2def}]), is examined in Appendix~\ref{app:referee}. We repeat te analysis considering only photometric errors, constant mass-to-light ratios, and  rearranging the true surface density profile to force a monotonic decrease of $\Sigma(R)$ in the inner region. None of these modifications alter than main conclusion that the 
resulting fits are quite good for a Schuster-Plummer potential and outrageous for a NFW potential.

One final outcome of the analysis is the radial extension of the potential parameterized by the characteristic radius $r_{sp}$. Considering the MCMC exploration of the posterior, the range of values is $\log(r_{sp}/1\,{\rm kpc})\simeq 0.79\pm 0.12$ for the Schuster-Plummer potential, $0.47 \pm 0.17$ for the NFW potential, and  $0.54 \pm 0.18$ for the $\rho_{230}$ potential. We note that their values cannot be compared directly since they are just scaling factors of different functional forms. One can compare the different potentials through the surface density giving rise to them. This comparison is included in Fig.~\ref{fig:df4_run_plotd} which, together with the profile of \nube\ , shows the mass surface density profiles corresponding to the three alternative potentials; the Schuster-Plummer potential (the blue line), the NFW potential (the green line), and the $\rho_{230}$ potential (the magenta line). Even though the spatial scaling of the potentials is set by the fit, the vertical scaling (i.e., the depth of the potential or its total mass) is arbitrary, and it was arbitrarily chosen in Fig.~\ref{fig:df4_run_plotd} to yield a stellar mass 30 times the stellar mass of \nube , which has $M_\star\simeq 3.9\times 10^8\,M_\odot$ as inferred from the best fitting profiles.    
\begin{figure}
\centering
\includegraphics[width=1.0\linewidth]{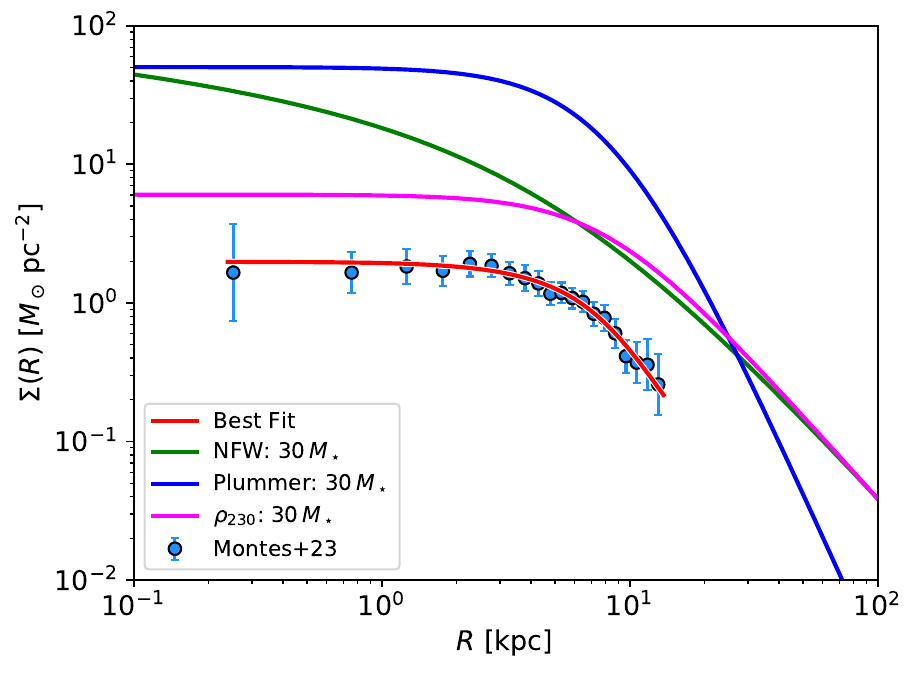} 
\caption{Comparison between the stellar surface density profile of \nube\  and the best-fitting potentials with $f\geq 0$. The plot shows the mass surface density that gives rise to the best-fitting Schuster-Plummer potential (blue line), NFW potential (green line), and $\rho_{230}$ potential (magenta line). The spatial scaling of the potentials is set by the fit whereas the vertical scaling is arbitrary, and it was arbitrarily chosen to be 30 times the stellar mass of \nube .
}
\label{fig:df4_run_plotd}
\end{figure}

We can also compare the observed density profile and the potentials using the core radius. Defined as the radius where the surface density is 1/2 the maximum value, $\Sigma(R_c)=\Sigma(0)/2$, it yields 
\def\caca{$\rho_{230} $} 
\begin{equation}
  \log\left[R_{cp}/R_{c}\right]=
\begin{cases}
0.06\pm 0.08 & {\rm Schuster-Plummer}, \\
 -0.01 \pm 0.13 & ~~~~~~~~~~~~~\rho_{230},
\end{cases}
  \label{eq:tricks}
\end{equation}
where $R_{c}$ and $R_{cp}$ stand for the core radius of the stars and the corresponding potential, respectively. (The NFW potential does not show a core and therefore is not included in Eq.~[\ref{eq:tricks}].)  The error bars comes from the standard deviation of the MCMC sampling of the posterior. We note that the core radius of stars and potential are similar.

%
\section{Simple extensions of the present formalism for anisotropic velocities}\label{sec:extensions}
Even though they are not used in the present work, the formalism for isotropic velocity distributions detailed in Sect.~\ref{sec:derivation} still holds, mutatis mutandis, for two  particularly interesting anisotropic velocity distribution. In the so-called  Osipkov-Merrit model, the velocity is assumed to have an anisotropy similar to the one expected in isolated dwarfs, with the orbits nearly isotropic in the center and tending to be radial in the outskirts \citep[e.g.,][]{2017ApJ...835..193E,2023MNRAS.525.3516O}. In this case one can write down \citep[e.g.,][]{2008gady.book.....B} an equation formally identical to Eq.~(\ref{eq:leading}) replacing the energy $\epsilon$ with $Q$,
\begin{equation}
  Q = \epsilon -\frac{L^2}{2r_b^2},
\end{equation}
and the volume density $\rho(r)$ with
\begin{equation}
\rho_{\rm OM}(r)= \left(1+\frac{r^2}{r_b^2}\right)\,\rho(r), 
\end{equation}
where $L$ stands for the norm of the angular momentum vector and $r_b$ is the characteristic radius where the transition between isotropic orbits (center) and radial orbits (outskirts) occurs. Thus, the tool developed in Sect.~\ref{sec:the_actual_algorithm} can be applied directly to $\rho_{\rm OM}$ to retrieve $f(Q)$. It would need to assume a value for $r_b$ but, given the speed of the fitting procedure, one can easily treat it as an hyper-parameter and derive $f(Q)$ given $r_b$.

Something similar happens with the case of an arbitrary but constant velocity anisotropy $\beta_u$ (defined in Eq.~[\ref{eq:ani-param}]). In this other case, the DF can be split as
\begin{equation}
f(\epsilon) = L^{-2\beta_u}\,f_\epsilon(\epsilon),
\end{equation}
and Eq.~(\ref{eq:leading}) can be replaced with \citep[e.g.,][]{2008gady.book.....B,2023ApJ...954..153S},
\begin{equation}
  r^{2\beta_u}\rho(r) = \kappa \,\int_0^{\Psi(r)}\frac{f_\epsilon(\epsilon)}{\left[\Psi(r)-\epsilon\right]^{\beta_u-1/2}}\,d\epsilon,
  \label{eq:th1}
\end{equation}
where $\kappa$ is a positive numerical value independent of the radius.  Note that this equation is formally quite similar to Eq.~(\ref{eq:leading}) provided $\beta_u < 1/2$, and so can be used to constraint $f(\epsilon)$ assuming a value for $\beta_u < 1/2$. Obviously, the equivalent to the characteristic densities (Eq.~[\ref{eq:master_mind}]) would have to be re-computed according to $\beta_u$, but preparing a battery of these functions for different $\beta_u$ is doable.

\section{Conclusions}\label{sec:conclusions}

We present a new technique to constrain some properties of the gravitational potential of a galaxy only from photometry, i.e., only from the distribution of stellar mass inferred from the observed starlight. Under a number of simplifying assumptions (spherical symmetry, isotropic velocity distribution, identical stars, and stationarity), the classical EIM (Sect.~\ref{sec:derivation}) allows us to infer the DF in the phase space needed for the observed stars to reside in an assumed gravitational potential. Thus, gravitational potential and starlight can be shown to be inconsistent if the required DF is negative somewhere. This seemingly simple idea has a far-reaching diagnostic capability in the context of understanding the nature of DM. The gravitational potential expected from CDM
is inconsistent with the central plateau or core often observed in the starlight distribution of dwarf galaxies  (Sect.~\ref{sec:intro}).

This new technique allowed \citet{2024arXiv240716755S} to point out possible deviations from the CDM paradigm. The implementation as a specific tool is detailed in Sect.~\ref{sec:derivation}, with several consistency tests also collected in Appendix~\ref{sec:sanity}. Sections \ref{sec:derivation}, \ref{sec:eigen_sp}, and \ref{sec:eigen_nfw} spell out the mathematical formulation whereas Sect.~\ref{sec:the_actual_algorithm} describes the actual numerical implementation of the general technique using a Bayesian approach.   
The low-surface brightness dwarf galaxy \nube\ recently discovered by \citet{2024A&A...681A..15M} has, among other properties,  a conspicuous and large inner core (Fig.~\ref{fig:nube} and Sect.~\ref{sec:observation}) used by \citeauthor{2024A&A...681A..15M} to work out the constraints on fuzzy DM imposed by the existence of such a core. \nube\ is used in this paper to showcase the application of our tool and to illustrate the kind of physical information it provides.

The actual application to \nube\ is described in Sect.~\ref{sec:potential_nube}. Provided \nube\ complies with the assumptions underlying EMI, cuspy NFW potentials are strongly disfavored compared to those with cores (Schuster-Plummer or $\rho_{230}$). As we explain in Sect.~\ref{sec:potential_nube_analytic}, the mild inner positive slope of \nube\ (Fig.~\ref{fig:nube} and Eq.~[\ref{eq:innermost}]) cannot be reproduced with $f(\epsilon)\geq 0$. However, when $f(\epsilon)$ is forced to be positive, the resulting fits assuming a core potential have smaller $\chi^2$ and innermost slopes closer to zero than the fits assuming NFW potentials. According to the statistical tests carried out in Sect.~\ref{sec:potential_nube}, the probability that the best Schuster-Plummer $f(\epsilon) \geq  0$ fit is better than the best NFW $f(\epsilon)\geq 0$ fit is around 89\,\%.
The fact that \nube\ resides in a potential that is not cuspy is not fully unexpected.  Its stellar mass,  $\sim3.9\times 10^8\,M_\odot$, is large enough for the baryon feedback to modify the inner region of the global potential, turning cusps into cores (Sect.~\ref{sec:intro}). The large size of core is surprising, though, a fact difficult to explain by the current cosmological CDM numerical simulations of low surface brightness galaxies \citep{2024A&A...681A..15M}. Another potential possibility to explain why \nube\ lives in a cored potential would be that the stars contribute significantly to the total mass of the system, so that the overall gravitational potential automatically follows the stellar distribution. However, this explanation is unlikely since the DM content estimated by \citet{2024A&A...681A..15M} is from 20 to 120 times larger than the stellar mass. We also studied the extent to which these conclusions depend on the estimated errors in the observed profile and the assumed mass-to-light ratio (Appendix~\ref{app:referee}). Neither of these two issues compromises the conclusion that cuspy profiles are disfavored.

The tool also allows us to  constrain the length scale of the potential (Eq.~[\ref{eq:tricks}]). In terms of the  core radius (i.e., when the density drops to half the central value), we find the cored potentials to be similar to the large stellar core shown by \nube\ (effective radius of $6.9\,{\rm kpc}$). The possibility of constraining the relation between the core size of stars and DM happens to be one of the interesting outcomes of EIM \citep[][]{2023ApJ...954..153S}.  

The EIM-based tool described in the paper has considerable room for improvement. The use of other potentials to represent the DM distribution is as simple as computing the required characteristic densities (Eq.~[\ref{eq:master_mind}]). Section~\ref{sec:extensions} sketches simple extensions that relax the need of isotropic velocities, so that the same kind of tool should work for systems with constant but anisotropic velocities and systems with gradients of anisotropy, from isotropic in their center to radially biased in the outskirts (aka Osipkov-Merrit). We note that  quasi-isotropic orbits  and Osipkov-Merrit-like velocity anisotropies are indeed preferred by the model dwarf galaxies formed in realistic cosmological numerical simulations \citep{2017ApJ...835..193E,2017MNRAS.472.4786G,2023MNRAS.525.3516O} and is also  found in dwarf spheroidal galaxies with observed kinematics \citep{2020A&A...633A..36M,2022A&A...659A.119K}. Moreover, \citet{2024arXiv240716519S} showed how stellar cores are also inconsistent with NFW potentials in axi-symmetric systems using an extension of the original EIM. This extension represents a solid starting point to develop the tool further, so that we can constrain the gravitational potential dropping the spherical symmetry assumption.


\begin{acknowledgements}
Thanks are due to Andr\'es Asensio for guiding us on the use of the Bayesian tools.
Thanks are also due to Ignacio Ferreras for insightful discussions on how to quantify the goodness of fits based on the two competing potentials.
JSA acknowledges financial support from the Spanish Ministry of Science and Innovation, project PID2022-136598NB-C31 (ESTALLIDOS8), and from Gobierno de Canarias through EU FEDER funding, project PID2020010050. His visit to La Plata was partly covered by the MICINN through the Spanish State Research Agency, under Severo Ochoa Centers of Excellence Programme 2020-2023 (CEX2019- 000920-S).
IT acknowledges support from the ACIISI, Consejer\'{i}a de Econom\'{i}a, Conocimiento y Empleo del Gobierno de Canarias and the European Regional Development Fund (ERDF) under grant with reference PROID2021010044 and from the State Research Agency (AEI-MCINN) of the Spanish Ministry of Science and Innovation under the grant PID2022-140869NB-I00 and IAC project P/302302, financed by the Ministry of Science and Innovation, through the State Budget and by the Canary Islands Department of Economy, Knowledge and Employment, through the Regional Budget of the Autonomous Community. MM acknowledges support from the project RYC2022-036949-I financed by the MICIU/AEI/10.13039/501100011033 and by Fondo Social Europeo Plus (FSE+). 
We acknowledge the use of the Python packages {\em numpy} \citep{2020Natur.585..357H}, {\em scipy} \citep{2020NatMe..17..261V}, {\em matplotlib} \citep{Hunter:2007}, and {\em emcee} \citep[][]{2013PASP..125..306F}. 
\end{acknowledgements}

%
%

%


\begin{thebibliography}{}
\expandafter\ifx\csname natexlab\endcsname\relax\def\natexlab#1{#1}\fi
\providecommand{\url}[1]{\href{#1}{#1}}
\providecommand{\dodoi}[1]{doi:~\href{http://doi.org/#1}{\nolinkurl{#1}}}
\providecommand{\doeprint}[1]{\href{http://ascl.net/#1}{\nolinkurl{http://ascl.net/#1}}}
\providecommand{\doarXiv}[1]{\href{https://arxiv.org/abs/#1}{\nolinkurl{https://arxiv.org/abs/#1}}}

\bibitem[{{An} \& {Zhao}(2013)}]{2013MNRAS.428.2805A}
{An}, J., \& {Zhao}, H. 2013, \mnras, 428, 2805, \dodoi{10.1093/mnras/sts175}

\bibitem[{{An} \& {Evans}(2006)}]{2006ApJ...642..752A}
{An}, J.~H., \& {Evans}, N.~W. 2006, \apj, 642, 752, \dodoi{10.1086/501040}

\bibitem[{{Battaglia} \& {Nipoti}(2022)}]{2022NatAs...6..659B}
{Battaglia}, G., \& {Nipoti}, C. 2022, Nature Astronomy, 6, 659,
  \dodoi{10.1038/s41550-022-01638-7}

\bibitem[{{Bechtol} {et~al.}(2022){Bechtol}, {Birrer}, {Cyr-Racine}, {Schutz},
  {Adhikari}, {Amin}, {Banerjee}, {Bird}, {Blinov}, {Boddy}, {Boehm}, {Bundy},
  {Buschmann}, {Chakrabarti}, {Curtin}, {Dai}, {Drlica-Wagner}, {Dvorkin},
  {Erickcek}, {Gilman}, {Heeba}, {Kim}, {Ir{\v{s}}i{\v{c}}}, {Leauthaud},
  {Lovell}, {Luki{\'c}}, {Mao}, {Mau}, {Mitridate}, {Mocz}, {Mu{\~n}oz},
  {Nadler}, {Peter}, {Price-Whelan}, {Robertson}, {Sabti}, {Sehgal}, {Shipp},
  {Simon}, {Singh}, {Van Tilburg}, {Wechsler}, {Widmark}, \&
  {Yu}}]{2022arXiv220307354B}
{Bechtol}, K., {Birrer}, S., {Cyr-Racine}, F.-Y., {et~al.} 2022, arXiv
  e-prints, arXiv:2203.07354, \dodoi{10.48550/arXiv.2203.07354}

\bibitem[{{Binney} \& {Tremaine}(2008)}]{2008gady.book.....B}
{Binney}, J., \& {Tremaine}, S. 2008, {Galactic Dynamics: Second Edition}

\bibitem[{{Bullock} \& {Boylan-Kolchin}(2017)}]{2017ARA&A..55..343B}
{Bullock}, J.~S., \& {Boylan-Kolchin}, M. 2017, \araa, 55, 343,
  \dodoi{10.1146/annurev-astro-091916-055313}

\bibitem[{{Carlsten} {et~al.}(2021){Carlsten}, {Greene}, {Greco}, {Beaton}, \&
  {Kado-Fong}}]{2021ApJ...922..267C}
{Carlsten}, S.~G., {Greene}, J.~E., {Greco}, J.~P., {Beaton}, R.~L., \&
  {Kado-Fong}, E. 2021, \apj, 922, 267, \dodoi{10.3847/1538-4357/ac2581}

\bibitem[{{Carr} {et~al.}(2024){Carr}, {Clesse}, {Garc{\'\i}a-Bellido},
  {Hawkins}, \& {K{\"u}hnel}}]{2024PhR..1054....1C}
{Carr}, B.~J., {Clesse}, S., {Garc{\'\i}a-Bellido}, J., {Hawkins}, M.~R.~S., \&
  {K{\"u}hnel}, F. 2024, \physrep, 1054, 1,
  \dodoi{10.1016/j.physrep.2023.11.005}

\bibitem[{{Chabrier}(2003)}]{2003ApJ...586L.133C}
{Chabrier}, G. 2003, \apjl, 586, L133, \dodoi{10.1086/374879}

\bibitem[{{Chan} {et~al.}(2015){Chan}, {Kere{\v{s}}}, {O{\~n}orbe}, {Hopkins},
  {Muratov}, {Faucher-Gigu{\`e}re}, \& {Quataert}}]{2015MNRAS.454.2981C}
{Chan}, T.~K., {Kere{\v{s}}}, D., {O{\~n}orbe}, J., {et~al.} 2015, \mnras, 454,
  2981, \dodoi{10.1093/mnras/stv2165}

\bibitem[{{Ciotti}(2021)}]{2021isd..book.....C}
{Ciotti}, L. 2021, {Introduction to Stellar Dynamics},
  \dodoi{10.1017/9780511736117}

\bibitem[{{Ciotti} \& {Morganti}(2010)}]{2010MNRAS.401.1091C}
{Ciotti}, L., \& {Morganti}, L. 2010, \mnras, 401, 1091,
  \dodoi{10.1111/j.1365-2966.2009.15697.x}

\bibitem[{{Del Popolo} \& {Le Delliou}(2017)}]{2017Galax...5...17D}
{Del Popolo}, A., \& {Le Delliou}, M. 2017, Galaxies, 5, 17,
  \dodoi{10.3390/galaxies5010017}

\bibitem[{{Dhillon} {et~al.}(2018){Dhillon}, {Dixon}, {Gamble}, {Kerry},
  {Littlefair}, {Parsons}, {Marsh}, {Bezawada}, {Black}, {Gao}, {Henry},
  {Lunney}, {Miller}, {Dubbeldam}, {Morris}, {Osborn}, {Wilson}, {Casares},
  {Mu{\~n}oz-Darias}, {Pall{\'e}}, {Rodriguez-Gil}, {Shahbaz}, \& {de Ugarte
  Postigo}}]{2018SPIE10702E..0LD}
{Dhillon}, V., {Dixon}, S., {Gamble}, T., {et~al.} 2018, in Society of
  Photo-Optical Instrumentation Engineers (SPIE) Conference Series, Vol. 10702,
  Ground-based and Airborne Instrumentation for Astronomy VII, ed. C.~J.
  {Evans}, L.~{Simard}, \& H.~{Takami}, 107020L, \dodoi{10.1117/12.2312041}

\bibitem[{{Dhillon} {et~al.}(2021){Dhillon}, {Bezawada}, {Black}, {Dixon},
  {Gamble}, {Gao}, {Henry}, {Kerry}, {Littlefair}, {Lunney}, {Marsh}, {Miller},
  {Parsons}, {Ashley}, {Breedt}, {Brown}, {Dyer}, {Green}, {Pelisoli},
  {Sahman}, {Wild}, {Ives}, {Mehrgan}, {Stegmeier}, {Dubbeldam}, {Morris},
  {Osborn}, {Wilson}, {Casares}, {Mu{\~n}oz-Darias}, {Pall{\'e}},
  {Rodr{\'\i}guez-Gil}, {Shahbaz}, {Torres}, {de Ugarte Postigo},
  {Cabrera-Lavers}, {Corradi}, {Dom{\'\i}nguez}, \&
  {Garc{\'\i}a-Alvarez}}]{2021MNRAS.507..350D}
{Dhillon}, V.~S., {Bezawada}, N., {Black}, M., {et~al.} 2021, \mnras, 507, 350,
  \dodoi{10.1093/mnras/stab2130}

\bibitem[{{Di Cintio} {et~al.}(2014{\natexlab{a}}){Di Cintio}, {Brook},
  {Dutton}, {Macci{\`o}}, {Stinson}, \& {Knebe}}]{2014MNRAS.441.2986D}
{Di Cintio}, A., {Brook}, C.~B., {Dutton}, A.~A., {et~al.} 2014{\natexlab{a}},
  \mnras, 441, 2986, \dodoi{10.1093/mnras/stu729}

\bibitem[{{Di Cintio} {et~al.}(2014{\natexlab{b}}){Di Cintio}, {Brook},
  {Macci{\`o}}, {Stinson}, {Knebe}, {Dutton}, \&
  {Wadsley}}]{2014MNRAS.437..415D}
{Di Cintio}, A., {Brook}, C.~B., {Macci{\`o}}, A.~V., {et~al.}
  2014{\natexlab{b}}, \mnras, 437, 415, \dodoi{10.1093/mnras/stt1891}

\bibitem[{{Dodelson} \& {Widrow}(1994)}]{1994PhRvL..72...17D}
{Dodelson}, S., \& {Widrow}, L.~M. 1994, \prl, 72, 17,
  \dodoi{10.1103/PhysRevLett.72.17}

\bibitem[{{Eddington}(1916)}]{1916MNRAS..76..572E}
{Eddington}, A.~S. 1916, \mnras, 76, 572, \dodoi{10.1093/mnras/76.7.572}

\bibitem[{{El-Badry} {et~al.}(2017){El-Badry}, {Wetzel}, {Geha}, {Quataert},
  {Hopkins}, {Kere{\v{s}}}, {Chan}, \&
  {Faucher-Gigu{\`e}re}}]{2017ApJ...835..193E}
{El-Badry}, K., {Wetzel}, A.~R., {Geha}, M., {et~al.} 2017, \apj, 835, 193,
  \dodoi{10.3847/1538-4357/835/2/193}

\bibitem[{{Foreman-Mackey} {et~al.}(2013){Foreman-Mackey}, {Hogg}, {Lang}, \&
  {Goodman}}]{2013PASP..125..306F}
{Foreman-Mackey}, D., {Hogg}, D.~W., {Lang}, D., \& {Goodman}, J. 2013, \pasp,
  125, 306, \dodoi{10.1086/670067}

\bibitem[{{Gonz{\'a}lez-Samaniego} {et~al.}(2017){Gonz{\'a}lez-Samaniego},
  {Bullock}, {Boylan-Kolchin}, {Fitts}, {Elbert}, {Hopkins}, {Kere{\v{s}}}, \&
  {Faucher-Gigu{\`e}re}}]{2017MNRAS.472.4786G}
{Gonz{\'a}lez-Samaniego}, A., {Bullock}, J.~S., {Boylan-Kolchin}, M., {et~al.}
  2017, \mnras, 472, 4786, \dodoi{10.1093/mnras/stx2322}

\bibitem[{{Governato} {et~al.}(2010){Governato}, {Brook}, {Mayer}, {Brooks},
  {Rhee}, {Wadsley}, {Jonsson}, {Willman}, {Stinson}, {Quinn}, \&
  {Madau}}]{2010Natur.463..203G}
{Governato}, F., {Brook}, C., {Mayer}, L., {et~al.} 2010, \nat, 463, 203,
  \dodoi{10.1038/nature08640}

\bibitem[{{Harris} {et~al.}(2020){Harris}, {Millman}, {van der Walt},
  {Gommers}, {Virtanen}, {Cournapeau}, {Wieser}, {Taylor}, {Berg}, {Smith},
  {Kern}, {Picus}, {Hoyer}, {van Kerkwijk}, {Brett}, {Haldane}, {del R{\'\i}o},
  {Wiebe}, {Peterson}, {G{\'e}rard-Marchant}, {Sheppard}, {Reddy}, {Weckesser},
  {Abbasi}, {Gohlke}, \& {Oliphant}}]{2020Natur.585..357H}
{Harris}, C.~R., {Millman}, K.~J., {van der Walt}, S.~J., {et~al.} 2020, \nat,
  585, 357, \dodoi{10.1038/s41586-020-2649-2}

\bibitem[{{Hayashi} {et~al.}(2020){Hayashi}, {Chiba}, \&
  {Ishiyama}}]{2020ApJ...904...45H}
{Hayashi}, K., {Chiba}, M., \& {Ishiyama}, T. 2020, \apj, 904, 45,
  \dodoi{10.3847/1538-4357/abbe0a}

\bibitem[{{Hernquist}(1990)}]{1990ApJ...356..359H}
{Hernquist}, L. 1990, \apj, 356, 359, \dodoi{10.1086/168845}

\bibitem[{{Hickstein} {et~al.}(2019){Hickstein}, {Gibson}, {Yurchak}, {Das}, \&
  {Ryazanov}}]{2019RScI...90f5115H}
{Hickstein}, D.~D., {Gibson}, S.~T., {Yurchak}, R., {Das}, D.~D., \&
  {Ryazanov}, M. 2019, Review of Scientific Instruments, 90, 065115,
  \dodoi{10.1063/1.5092635}

\bibitem[{{Hu} {et~al.}(2000){Hu}, {Barkana}, \&
  {Gruzinov}}]{2000PhRvL..85.1158H}
{Hu}, W., {Barkana}, R., \& {Gruzinov}, A. 2000, \prl, 85, 1158,
  \dodoi{10.1103/PhysRevLett.85.1158}

\bibitem[{Hunter(2007)}]{Hunter:2007}
Hunter, J.~D. 2007, Computing in Science \& Engineering, 9, 90,
  \dodoi{10.1109/MCSE.2007.55}

\bibitem[{{Ivezi{\'c}} {et~al.}(2019){Ivezi{\'c}}, {Kahn}, {Tyson}, {Abel},
  {Acosta}, {Allsman}, {Alonso}, {AlSayyad}, {Anderson}, {Andrew}, {Angel},
  {Angeli}, {Ansari}, {Antilogus}, {Araujo}, {Armstrong}, {Arndt}, {Astier},
  {Aubourg}, {Auza}, {Axelrod}, {Bard}, {Barr}, {Barrau}, {Bartlett}, {Bauer},
  {Bauman}, {Baumont}, {Bechtol}, {Bechtol}, {Becker}, {Becla}, {Beldica},
  {Bellavia}, {Bianco}, {Biswas}, {Blanc}, {Blazek}, {Blandford}, {Bloom},
  {Bogart}, {Bond}, {Booth}, {Borgland}, {Borne}, {Bosch}, {Boutigny},
  {Brackett}, {Bradshaw}, {Brandt}, {Brown}, {Bullock}, {Burchat}, {Burke},
  {Cagnoli}, {Calabrese}, {Callahan}, {Callen}, {Carlin}, {Carlson},
  {Chandrasekharan}, {Charles-Emerson}, {Chesley}, {Cheu}, {Chiang}, {Chiang},
  {Chirino}, {Chow}, {Ciardi}, {Claver}, {Cohen-Tanugi}, {Cockrum}, {Coles},
  {Connolly}, {Cook}, {Cooray}, {Covey}, {Cribbs}, {Cui}, {Cutri}, {Daly},
  {Daniel}, {Daruich}, {Daubard}, {Daues}, {Dawson}, {Delgado}, {Dellapenna},
  {de Peyster}, {de Val-Borro}, {Digel}, {Doherty}, {Dubois},
  {Dubois-Felsmann}, {Durech}, {Economou}, {Eifler}, {Eracleous}, {Emmons},
  {Fausti Neto}, {Ferguson}, {Figueroa}, {Fisher-Levine}, {Focke}, {Foss},
  {Frank}, {Freemon}, {Gangler}, {Gawiser}, {Geary}, {Gee}, {Geha}, {Gessner},
  {Gibson}, {Gilmore}, {Glanzman}, {Glick}, {Goldina}, {Goldstein}, {Goodenow},
  {Graham}, {Gressler}, {Gris}, {Guy}, {Guyonnet}, {Haller}, {Harris},
  {Hascall}, {Haupt}, {Hernandez}, {Herrmann}, {Hileman}, {Hoblitt}, {Hodgson},
  {Hogan}, {Howard}, {Huang}, {Huffer}, {Ingraham}, {Innes}, {Jacoby}, {Jain},
  {Jammes}, {Jee}, {Jenness}, {Jernigan}, {Jevremovi{\'c}}, {Johns}, {Johnson},
  {Johnson}, {Jones}, {Juramy-Gilles}, {Juri{\'c}}, {Kalirai}, {Kallivayalil},
  {Kalmbach}, {Kantor}, {Karst}, {Kasliwal}, {Kelly}, {Kessler}, {Kinnison},
  {Kirkby}, {Knox}, {Kotov}, {Krabbendam}, {Krughoff}, {Kub{\'a}nek},
  {Kuczewski}, {Kulkarni}, {Ku}, {Kurita}, {Lage}, {Lambert}, {Lange},
  {Langton}, {Le Guillou}, {Levine}, {Liang}, {Lim}, {Lintott}, {Long},
  {Lopez}, {Lotz}, {Lupton}, {Lust}, {MacArthur}, {Mahabal}, {Mandelbaum},
  {Markiewicz}, {Marsh}, {Marshall}, {Marshall}, {May}, {McKercher}, {McQueen},
  {Meyers}, {Migliore}, {Miller}, {Mills}, {Miraval}, {Moeyens}, {Moolekamp},
  {Monet}, {Moniez}, {Monkewitz}, {Montgomery}, {Morrison}, {Mueller},
  {Muller}, {Mu{\~n}oz Arancibia}, {Neill}, {Newbry}, {Nief}, {Nomerotski},
  {Nordby}, {O'Connor}, {Oliver}, {Olivier}, {Olsen}, {O'Mullane}, {Ortiz},
  {Osier}, {Owen}, {Pain}, {Palecek}, {Parejko}, {Parsons}, {Pease},
  {Peterson}, {Peterson}, {Petravick}, {Libby Petrick}, {Petry},
  {Pierfederici}, {Pietrowicz}, {Pike}, {Pinto}, {Plante}, {Plate}, {Plutchak},
  {Price}, {Prouza}, {Radeka}, {Rajagopal}, {Rasmussen}, {Regnault}, {Reil},
  {Reiss}, {Reuter}, {Ridgway}, {Riot}, {Ritz}, {Robinson}, {Roby}, {Roodman},
  {Rosing}, {Roucelle}, {Rumore}, {Russo}, {Saha}, {Sassolas}, {Schalk},
  {Schellart}, {Schindler}, {Schmidt}, {Schneider}, {Schneider}, {Schoening},
  {Schumacher}, {Schwamb}, {Sebag}, {Selvy}, {Sembroski}, {Seppala}, {Serio},
  {Serrano}, {Shaw}, {Shipsey}, {Sick}, {Silvestri}, {Slater}, {Smith},
  {Smith}, {Sobhani}, {Soldahl}, {Storrie-Lombardi}, {Stover}, {Strauss},
  {Street}, {Stubbs}, {Sullivan}, {Sweeney}, {Swinbank}, {Szalay}, {Takacs},
  {Tether}, {Thaler}, {Thayer}, {Thomas}, {Thornton}, {Thukral}, {Tice},
  {Trilling}, {Turri}, {Van Berg}, {Vanden Berk}, {Vetter}, {Virieux},
  {Vucina}, {Wahl}, {Walkowicz}, {Walsh}, {Walter}, {Wang}, {Wang}, {Warner},
  {Wiecha}, {Willman}, {Winters}, {Wittman}, {Wolff}, {Wood-Vasey}, {Wu},
  {Xin}, {Yoachim}, \& {Zhan}}]{2019ApJ...873..111I}
{Ivezi{\'c}}, {\v{Z}}., {Kahn}, S.~M., {Tyson}, J.~A., {et~al.} 2019, \apj,
  873, 111, \dodoi{10.3847/1538-4357/ab042c}

\bibitem[{{Jackson} {et~al.}(2021){Jackson}, {Martin}, {Kaviraj}, {Rams{\o}y},
  {Devriendt}, {Sedgwick}, {Laigle}, {Choi}, {Beckmann}, {Volonteri}, {Dubois},
  {Pichon}, {Yi}, {Slyz}, {Kraljic}, {Kimm}, {Peirani}, \&
  {Baldry}}]{2021MNRAS.502.4262J}
{Jackson}, R.~A., {Martin}, G., {Kaviraj}, S., {et~al.} 2021, \mnras, 502,
  4262, \dodoi{10.1093/mnras/stab077}

\bibitem[{{Koudmani} {et~al.}(2024){Koudmani}, {Rennehan}, {Somerville},
  {Hayward}, {Angl{\'e}s-Alc{\'a}zar}, {Orr}, {Sands}, \&
  {Wellons}}]{2024arXiv240902172K}
{Koudmani}, S., {Rennehan}, D., {Somerville}, R.~S., {et~al.} 2024, arXiv
  e-prints, arXiv:2409.02172, \dodoi{10.48550/arXiv.2409.02172}

\bibitem[{{Kowalczyk} \& {{\L}okas}(2022)}]{2022A&A...659A.119K}
{Kowalczyk}, K., \& {{\L}okas}, E.~L. 2022, \aap, 659, A119,
  \dodoi{10.1051/0004-6361/202142212}

\bibitem[{{Lacroix} {et~al.}(2018){Lacroix}, {Stref}, \&
  {Lavalle}}]{2018JCAP...09..040L}
{Lacroix}, T., {Stref}, M., \& {Lavalle}, J. 2018, \jcap, 2018, 040,
  \dodoi{10.1088/1475-7516/2018/09/040}

\bibitem[{{Laureijs} {et~al.}(2011){Laureijs}, {Amiaux}, {Arduini},
  {Augu{\`e}res}, {Brinchmann}, {Cole}, {Cropper}, {Dabin}, {Duvet}, {Ealet},
  {Garilli}, {Gondoin}, {Guzzo}, {Hoar}, {Hoekstra}, {Holmes}, {Kitching},
  {Maciaszek}, {Mellier}, {Pasian}, {Percival}, {Rhodes}, {Saavedra Criado},
  {Sauvage}, {Scaramella}, {Valenziano}, {Warren}, {Bender}, {Castander},
  {Cimatti}, {Le F{\`e}vre}, {Kurki-Suonio}, {Levi}, {Lilje}, {Meylan},
  {Nichol}, {Pedersen}, {Popa}, {Rebolo Lopez}, {Rix}, {Rottgering},
  {Zeilinger}, {Grupp}, {Hudelot}, {Massey}, {Meneghetti}, {Miller}, {Paltani},
  {Paulin-Henriksson}, {Pires}, {Saxton}, {Schrabback}, {Seidel}, {Walsh},
  {Aghanim}, {Amendola}, {Bartlett}, {Baccigalupi}, {Beaulieu}, {Benabed},
  {Cuby}, {Elbaz}, {Fosalba}, {Gavazzi}, {Helmi}, {Hook}, {Irwin}, {Kneib},
  {Kunz}, {Mannucci}, {Moscardini}, {Tao}, {Teyssier}, {Weller}, {Zamorani},
  {Zapatero Osorio}, {Boulade}, {Foumond}, {Di Giorgio}, {Guttridge}, {James},
  {Kemp}, {Martignac}, {Spencer}, {Walton}, {Bl{\"u}mchen}, {Bonoli},
  {Bortoletto}, {Cerna}, {Corcione}, {Fabron}, {Jahnke}, {Ligori}, {Madrid},
  {Martin}, {Morgante}, {Pamplona}, {Prieto}, {Riva}, {Toledo}, {Trifoglio},
  {Zerbi}, {Abdalla}, {Douspis}, {Grenet}, {Borgani}, {Bouwens}, {Courbin},
  {Delouis}, {Dubath}, {Fontana}, {Frailis}, {Grazian}, {Koppenh{\"o}fer},
  {Mansutti}, {Melchior}, {Mignoli}, {Mohr}, {Neissner}, {Noddle}, {Poncet},
  {Scodeggio}, {Serrano}, {Shane}, {Starck}, {Surace}, {Taylor},
  {Verdoes-Kleijn}, {Vuerli}, {Williams}, {Zacchei}, {Altieri}, {Escudero
  Sanz}, {Kohley}, {Oosterbroek}, {Astier}, {Bacon}, {Bardelli}, {Baugh},
  {Bellagamba}, {Benoist}, {Bianchi}, {Biviano}, {Branchini}, {Carbone},
  {Cardone}, {Clements}, {Colombi}, {Conselice}, {Cresci}, {Deacon}, {Dunlop},
  {Fedeli}, {Fontanot}, {Franzetti}, {Giocoli}, {Garcia-Bellido}, {Gow},
  {Heavens}, {Hewett}, {Heymans}, {Holland}, {Huang}, {Ilbert}, {Joachimi},
  {Jennins}, {Kerins}, {Kiessling}, {Kirk}, {Kotak}, {Krause}, {Lahav}, {van
  Leeuwen}, {Lesgourgues}, {Lombardi}, {Magliocchetti}, {Maguire}, {Majerotto},
  {Maoli}, {Marulli}, {Maurogordato}, {McCracken}, {McLure}, {Melchiorri},
  {Merson}, {Moresco}, {Nonino}, {Norberg}, {Peacock}, {Pello}, {Penny},
  {Pettorino}, {Di Porto}, {Pozzetti}, {Quercellini}, {Radovich}, {Rassat},
  {Roche}, {Ronayette}, {Rossetti}, {Sartoris}, {Schneider}, {Semboloni},
  {Serjeant}, {Simpson}, {Skordis}, {Smadja}, {Smartt}, {Spano}, {Spiro},
  {Sullivan}, {Tilquin}, {Trotta}, {Verde}, {Wang}, {Williger}, {Zhao},
  {Zoubian}, \& {Zucca}}]{2011arXiv1110.3193L}
{Laureijs}, R., {Amiaux}, J., {Arduini}, S., {et~al.} 2011, arXiv e-prints,
  arXiv:1110.3193, \dodoi{10.48550/arXiv.1110.3193}

\bibitem[{{Massari} {et~al.}(2020){Massari}, {Helmi}, {Mucciarelli}, {Sales},
  {Spina}, \& {Tolstoy}}]{2020A&A...633A..36M}
{Massari}, D., {Helmi}, A., {Mucciarelli}, A., {et~al.} 2020, \aap, 633, A36,
  \dodoi{10.1051/0004-6361/201935613}

\bibitem[{{Merritt} {et~al.}(2006){Merritt}, {Graham}, {Moore}, {Diemand}, \&
  {Terzi{\'c}}}]{2006AJ....132.2685M}
{Merritt}, D., {Graham}, A.~W., {Moore}, B., {Diemand}, J., \& {Terzi{\'c}}, B.
  2006, \aj, 132, 2685, \dodoi{10.1086/508988}

\bibitem[{{Montes} {et~al.}(2024){Montes}, {Trujillo}, {Karunakaran},
  {Infante-Sainz}, {Spekkens}, {Golini}, {Beasley}, {Cebri{\'a}n}, {Chamba},
  {D'Onofrio}, {Kelvin}, \& {Rom{\'a}n}}]{2024A&A...681A..15M}
{Montes}, M., {Trujillo}, I., {Karunakaran}, A., {et~al.} 2024, \aap, 681, A15,
  \dodoi{10.1051/0004-6361/202347667}

\bibitem[{{Moskowitz} \& {Walker}(2020)}]{2020ApJ...892...27M}
{Moskowitz}, A.~G., \& {Walker}, M.~G. 2020, \apj, 892, 27,
  \dodoi{10.3847/1538-4357/ab7459}

\bibitem[{{Navarro} {et~al.}(1997){Navarro}, {Frenk}, \&
  {White}}]{1997ApJ...490..493N}
{Navarro}, J.~F., {Frenk}, C.~S., \& {White}, S. D.~M. 1997, \apj, 490, 493,
  \dodoi{10.1086/304888}

\bibitem[{{Orkney} {et~al.}(2023){Orkney}, {Taylor}, {Read}, {Rey}, {Pontzen},
  {Agertz}, {Kim}, \& {Delorme}}]{2023MNRAS.525.3516O}
{Orkney}, M. D.~A., {Taylor}, E., {Read}, J.~I., {et~al.} 2023, \mnras, 525,
  3516, \dodoi{10.1093/mnras/stad2516}

\bibitem[{{Pe{\~n}arrubia} {et~al.}(2012){Pe{\~n}arrubia}, {Pontzen}, {Walker},
  \& {Koposov}}]{2012ApJ...759L..42P}
{Pe{\~n}arrubia}, J., {Pontzen}, A., {Walker}, M.~G., \& {Koposov}, S.~E. 2012,
  \apjl, 759, L42, \dodoi{10.1088/2041-8205/759/2/L42}

\bibitem[{{Plastino} \& {Plastino}(1993)}]{1993PhLA..174..384P}
{Plastino}, A.~R., \& {Plastino}, A. 1993, Physics Letters A, 174, 384,
  \dodoi{10.1016/0375-9601(93)90195-6}

\bibitem[{{Pontzen} \& {Governato}(2012)}]{2012MNRAS.421.3464P}
{Pontzen}, A., \& {Governato}, F. 2012, \mnras, 421, 3464,
  \dodoi{10.1111/j.1365-2966.2012.20571.x}

\bibitem[{{Read} {et~al.}(2016){Read}, {Agertz}, \&
  {Collins}}]{2016MNRAS.459.2573R}
{Read}, J.~I., {Agertz}, O., \& {Collins}, M.~L.~M. 2016, \mnras, 459, 2573,
  \dodoi{10.1093/mnras/stw713}

\bibitem[{{Richstein} {et~al.}(2024){Richstein}, {Kallivayalil}, {Simon},
  {Garling}, {Wetzel}, {Warfield}, {van der Marel}, {Jeon}, {Rose}, {Torrey},
  {Engelhardt}, {Besla}, {Choi}, {Geha}, {Guhathakurta}, {Kirby}, {Patel},
  {Sacchi}, \& {Sohn}}]{2024ApJ...967...72R}
{Richstein}, H., {Kallivayalil}, N., {Simon}, J.~D., {et~al.} 2024, \apj, 967,
  72, \dodoi{10.3847/1538-4357/ad393c}

\bibitem[{{Roediger} \& {Courteau}(2015)}]{2015MNRAS.452.3209R}
{Roediger}, J.~C., \& {Courteau}, S. 2015, \mnras, 452, 3209,
  \dodoi{10.1093/mnras/stv1499}

\bibitem[{{Salucci}(2019)}]{2019A&ARv..27....2S}
{Salucci}, P. 2019, \aapr, 27, 2, \dodoi{10.1007/s00159-018-0113-1}

\bibitem[{{S{\'a}nchez Almeida}(2022)}]{2022Univ....8..214S}
{S{\'a}nchez Almeida}, J. 2022, Universe, 8, 214,
  \dodoi{10.3390/universe8040214}

\bibitem[{{S{\'a}nchez Almeida}(2024)}]{2024RNAAS...8..167S}
---. 2024, Research Notes of the American Astronomical Society, 8, 167,
  \dodoi{10.3847/2515-5172/ad5a0f}

\bibitem[{{S{\'a}nchez Almeida} {et~al.}(2023){S{\'a}nchez Almeida},
  {Plastino}, \& {Trujillo}}]{2023ApJ...954..153S}
{S{\'a}nchez Almeida}, J., {Plastino}, A.~R., \& {Trujillo}, I. 2023, \apj,
  954, 153, \dodoi{10.3847/1538-4357/ace534}

\bibitem[{{Sanchez Almeida} {et~al.}(2024{\natexlab{a}}){Sanchez Almeida},
  {Plastino}, \& {Trujillo}}]{2024arXiv240716519S}
{Sanchez Almeida}, J., {Plastino}, A.~R., \& {Trujillo}, I. 2024{\natexlab{a}},
  arXiv e-prints, arXiv:2407.16519, \dodoi{10.48550/arXiv.2407.16519}

\bibitem[{{S{\'a}nchez Almeida} {et~al.}(2020){S{\'a}nchez Almeida},
  {Trujillo}, \& {Plastino}}]{2020A&A...642L..14S}
{S{\'a}nchez Almeida}, J., {Trujillo}, I., \& {Plastino}, A.~R. 2020, \aap,
  642, L14, \dodoi{10.1051/0004-6361/202039190}

\bibitem[{{S{\'a}nchez Almeida} {et~al.}(2021){S{\'a}nchez Almeida},
  {Trujillo}, \& {Plastino}}]{2021ApJ...921..125S}
---. 2021, \apj, 921, 125, \dodoi{10.3847/1538-4357/ac1ba8}

\bibitem[{{Sanchez Almeida} {et~al.}(2024{\natexlab{b}}){Sanchez Almeida},
  {Trujillo}, \& {Plastino}}]{2024arXiv240716755S}
{Sanchez Almeida}, J., {Trujillo}, I., \& {Plastino}, A.~R. 2024{\natexlab{b}},
  arXiv e-prints, arXiv:2407.16755, \dodoi{10.48550/arXiv.2407.16755}

\bibitem[{{Spergel} \& {Steinhardt}(2000)}]{2000PhRvL..84.3760S}
{Spergel}, D.~N., \& {Steinhardt}, P.~J. 2000, \prl, 84, 3760,
  \dodoi{10.1103/PhysRevLett.84.3760}

\bibitem[{{Trujillo} {et~al.}(2021){Trujillo}, {D'Onofrio}, {Zaritsky},
  {Madrigal-Aguado}, {Chamba}, {Golini}, {Akhlaghi}, {Sharbaf},
  {Infante-Sainz}, {Rom{\'a}n}, {Morales-Socorro}, {Sand}, \&
  {Martin}}]{2021A&A...654A..40T}
{Trujillo}, I., {D'Onofrio}, M., {Zaritsky}, D., {et~al.} 2021, \aap, 654, A40,
  \dodoi{10.1051/0004-6361/202141603}

\bibitem[{Virtanen {et~al.}(2020)Virtanen, Gommers, Oliphant, Haberland, Reddy,
  Cournapeau, Burovski, Peterson, Weckesser, Bright, {van der Walt}, Brett,
  Wilson, Millman, Mayorov, Nelson, Jones, Kern, Larson, Carey, Polat, Feng,
  Moore, {VanderPlas}, Laxalde, Perktold, Cimrman, Henriksen, Quintero, Harris,
  Archibald, Ribeiro, Pedregosa, {van Mulbregt}, \& {SciPy 1.0
  Contributors}}]{2020SciPy-NMeth}
Virtanen, P., Gommers, R., Oliphant, T.~E., {et~al.} 2020, Nature Methods, 17,
  261, \dodoi{10.1038/s41592-019-0686-2}

\bibitem[{{Virtanen} {et~al.}(2020){Virtanen}, {Gommers}, {Oliphant},
  {Haberland}, {Reddy}, {Cournapeau}, {Burovski}, {Peterson}, {Weckesser},
  {Bright}, {van der Walt}, {Brett}, {Wilson}, {Millman}, {Mayorov}, {Nelson},
  {Jones}, {Kern}, {Larson}, {Carey}, {Polat}, {Feng}, {Moore}, {VanderPlas},
  {Laxalde}, {Perktold}, {Cimrman}, {Henriksen}, {Quintero}, {Harris},
  {Archibald}, {Ribeiro}, {Pedregosa}, {van Mulbregt}, \& {SciPy 1. 0
  Contributors}}]{2020NatMe..17..261V}
{Virtanen}, P., {Gommers}, R., {Oliphant}, T.~E., {et~al.} 2020, Nature
  Methods, 17, 261, \dodoi{10.1038/s41592-019-0686-2}

\bibitem[{{Zaritsky} {et~al.}(2024){Zaritsky}, {Golini}, {Donnerstein},
  {Trujillo}, {Akhlaghi}, {Chamba}, {D'Onofrio}, {Eskandarlou},
  {Hosseini-ShahiSavandi}, {Infante-Sainz}, {Martin}, {Montes}, {Rom{\'a}n},
  {Sedighi}, \& {Sharbaf}}]{2024AJ....168...69Z}
{Zaritsky}, D., {Golini}, G., {Donnerstein}, R., {et~al.} 2024, \aj, 168, 69,
  \dodoi{10.3847/1538-3881/ad543f}

\end{thebibliography}

%

\appendix


\section{Effects on \nube\ of errors in determining the galaxy center}\label{app:error_center}
\begin{figure}
\centering
\includegraphics[width=\linewidth]{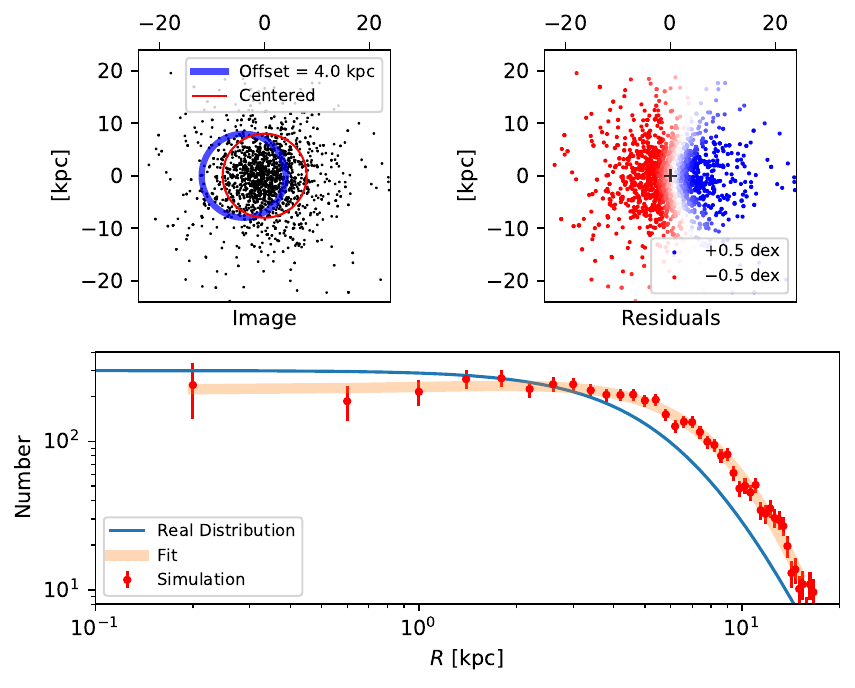}
\caption{MC simulation to show the effect of using a wrong center to compute the stellar mass surface density radial profile of a galaxy.
Top left panel: the dots represent individual stars. It shows only $10^3$ of them to avoid overcrowding but the actual simulation has $10^4$. The red circle is centered in the true distribution whereas the blue circle is offset by 4\,kpc. 
Top right panel: same as the left panel color coded with the residual left when subtracting a radial profile  computed with the 4\,kpc offset (the thick orange line in the bottom panel).   
Bottom panel: surface density profiles  (the symbols with error bars) computed using $\sim 50$ rings offset from the true center as indicated in the top left panel.  The error bars give the Poisson noise arising when counting stars. The true centered distribution is shown in blue.}
  \label{fig:off_center2a}
\end{figure}
This appendix studies whether the inner drop in the density profile of \nube\ (Fig.~\ref{fig:nube}) could be an artifact produced by selecting the wrong center when creating the surface density profile. To analyze the possibility, we carried out a series of Monte Carlo (MC) simulations like the one shown in Fig.~\ref{fig:off_center2a}.

Mock stars are randomly produced (the dots in the top left panel of Fig.~\ref{fig:off_center2a}) following a surface density mimicking \nube\ (the blue solid line in the bottom panel of Fig.~\ref{fig:off_center2a}). \nube\ is represented by a Schuster-Plummer profile (Eq.~\ref{eq:rhoabc}], where $a=2$, $b=5$, and $c = 0$) with a core radius of 8~kpc. Then the mock stars were counted in off-centered rings (e.g., the blue ring in Fig.~\ref{fig:off_center2a}, top left panel)  to produce density profiles like the red symbols with error bars in Fig.~\ref{fig:off_center2a}, bottom panel. (The error bars account for the Poisson error when counting.) Note that the resulting density profile shows a drop in the innermost regions of the simulated profile (Fig.~\ref{fig:off_center2a}, bottom panel, red symbols) which is not present in the original density profile (the blue solid line). A function like the one used to reproduce \nube\ (Eq.~[\ref{eq:rhoabc}]), which allows for a variable inner slope, was used to fit the mock density profile (the orange thick line in Fig.~\ref{fig:off_center2a}, bottom panel). The fitted profile has positive inner slope ($c < 0$, in the 3D profile). 

Simulations like the ones shown in Fig.~\ref{fig:off_center2a} were repeated many times to produce the summary plot represented in Fig.~\ref{fig:off_center5_plotc}. Each point is the mean of a 100 different realizations of a mock galaxy having the same parameters (i.e, same profile, same offset, and same number of stars), with the error bars showing the standard deviation among all these realizations. Different colors represent different number of stars in the mock galaxy, which is a proxy for the error bars in the radial profile. The profile shown in the bottom panel of Fig.~\ref{fig:off_center2a} corresponds to MC realizations with $10^4$ stars per galaxy. The effective noise with this number of stars is similar to the one shown by the observations of \nube\ (Fig.~\ref{fig:nube}) but, for the sake of  comprehensiveness, we also show values for galaxies simulated with $10^3$ stars (the orange symbols; too noisy for \nube) and with $10^5$ stars (the green symbols; too good for \nube). 
Figure~\ref{fig:off_center5_plotc} also includes a point representing the observation of \nube\ (the star symbol).  We have been very generous with the error bars assigned to \nube . The formal error bar assigned to its center by the algorithm to compute the surface density profile (Sect.~\ref{sec:observation}) is tiny ($\sim 0.6$\,pix, which render $\sim 0.025$\,kpc considering a plate scale of 0.08\,arcsec\,pix$^{-1}$ and a scale of 0.5\,kpc\,arcsec$^{-1}$).  Figure~\ref{fig:off_center5_plotc} includes error bars corresponding to 1/10 of the core radius, which is very conservative. As far as the inner slope, we use the mean of the two values found when the last point of the observed profile is or not included in the fit (see Fig.~\ref{fig:nube} and Sect.~\ref{sec:fit}), with the error bar being the semi-difference between the two.   
\begin{figure}
\centering
\includegraphics[width=\linewidth]{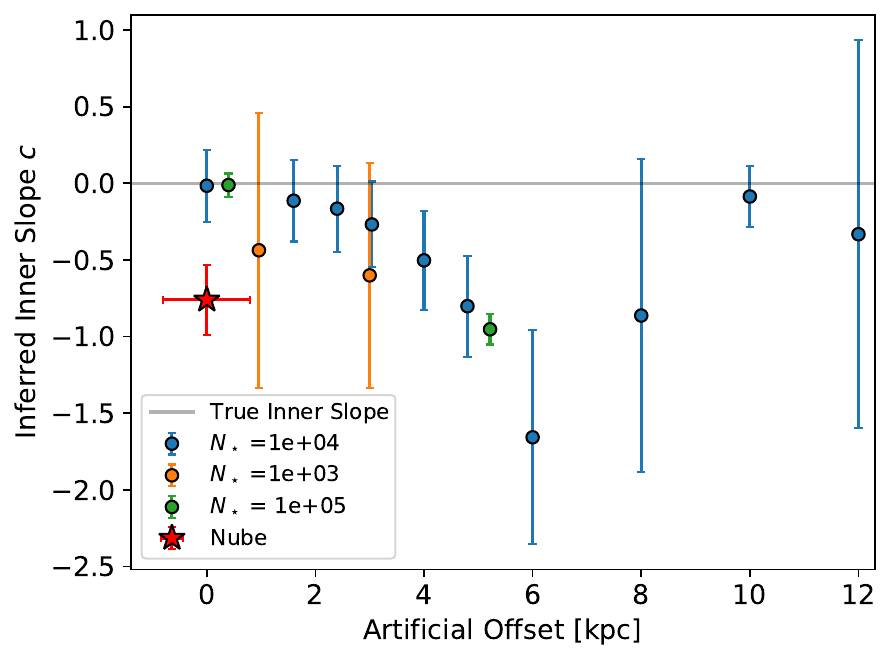}
\caption{Summary of the MC
to show the effect of using wrong centers to compute the stellar mass surface density radial profile of a galaxy. Inferred inner slope versus artificial offset (in kpc). The true inner slope is zero, as is the slope inferred from the fits when the offset goes to zero. The different colors represent different number of stars used to construct the mock galaxies (see the inset for the actual values). Those with errors closest to  \nube\ are represented with blue symbols. \nube\ is inconsistent with them.}
  \label{fig:off_center5_plotc}
\end{figure}

Given the MC simulations described above, a number of arguments discard that the negative inner slope of \nube\ is caused by an error in the center used to compute the density profile. (1)~The residuals of the fitting leave a dipole-like pattern (Fig.~\ref{fig:off_center2a}, top right panel) which is not the residual observed in \nube\ \citep[Fig.~11, right panel, in][]{2024A&A...681A..15M}. In other words, it is difficult to reconcile the residuals left by GALFIT on \nube\ with a significant global shift of the fitted function, in agreement with the negligible error that GALFIT provides ($\sim 0.025$\,kpc).
(2)~Considering the MC simulation that represents \nube\ best (blue symbols in Fig.~\ref{fig:off_center5_plotc}) one needs an artificial offset of some 5\,kpc to reproduce the observed slope, which is comparable with the core radius of \nube\ and, therefore, unrealistically large.
(3)~The results are robust in the sense that changing the hyper-parameters that define the MC simulation do not alter the conclusions (Appendix~\ref{sec:hyper_param}).
(4)~Other bias may create it (Appendix~\ref{app:referee}).
%
\section{Characteristic densities for the potential arising from $\rho_{\lowercase{abc}}$ profiles}\label{sec:rhoAbc}

As we did in \citet{2023ApJ...954..153S}, the case of a potential where the inner slope of the corresponding density is not zero (Schuster-Plummer) or minus one (NFW) can be treated in quite general terms using a $\rho_{abc}$ profile as defined in Eq.~(\ref{eq:rhoabc}). Using the Poisson equation for a spherically symmetric system \citep[e.g.,][]{2013MNRAS.428.2805A}, the potential is,
\begin{equation}
  \Psi_{abc}(r) =\frac{G\,M_{abc}(<r)}{r}+4\pi G\,\int_r^{\infty}\,t\,\rho_{abc}(t)\,dt,
  \label{eq:pot_general}
\end{equation}
with
\begin{displaymath}
  M_{abc}(<r) = 4\pi\,\int_0^r\,t^2\,\rho_{abc}(t)\,dt,
\end{displaymath}
so that 
\begin{equation}
  \Psi_{abc}(r)  = \epsilon_{max}  \frac{K(r/r_{sp},a,b,c)}{K(0,a,b,c)},
\end{equation}
with
\begin{equation}
\epsilon_{max}=  4\pi G\rho_{sp}r_{sp}^2\, K(0,a,b,c),
\end{equation}
and
\begin{equation}
K(x,a,b,c) =   \frac{1}{x}  \int^{x}_{0}\frac{t^{2-c}}{(1+t^a)^{(b-c)/a}}\,dt
+\int_{x}^{\infty}\frac{t^{1-c}}{(1+t^a)^{(b-c)/a}}\,dt.
\end{equation}
Then the characteristic density $\xi_{abc}$ follows from Eq.~(\ref{eq:master_mind}). We use this approach to carry out the fits for $\rho_{230}$ potentials described in the main text (e.g., Fig.~\ref{fig:df4_run_plot_3}). In this case, $a=2$, $b=3$, and $c=0$, which leads to a density with a core like a Schuster-Plummer profile and an outskirt similar to a NFW profile; see Fig.~\ref{fig:df4_run_plotd}.

\cprotect\section{Classical interpretation of Eqs.~(\ref{eq:mass_contri}) and~(\ref{eq:little_mass})}\label{app:classical}

  The variable $\varpi(\epsilon)$, defined in Eq.(\ref{eq:little_mass}) and appearing in Eq.(\ref{eq:mass_contri}), is interpreted in Sect.~\ref{sec:derivation} as the mass corresponding to each relative energy $\epsilon$. This variable, $\varpi(\epsilon)$, basically coincides with the quantity that in classical statistical mechanics is known as the {\em density of states}. It is usually denoted as $g(E)$, where, in the context of stellar dynamics, $E$ is the energy per unit mass
\citep[see][page 292, Eq.~(4.56)]{2008gady.book.....B}. The density of states $g(E)$ represents the phase-space volume per unit energy. That is, $g(E) dE$ is the volume in phase-space corresponding to particles with energies in the range $(E, E+dE)$. The quantity defined by us in Eq. (\ref{eq:little_mass}) represents the density of states expressed in terms of the relative energy. In fact, our Fig.~\ref{fig:mass_epsilon}, which depicts $\varpi(\epsilon)$ for various potentials relevant for the present work, looks, qualitatively, as a mirror image of Fig.~4.3 of \citet{2008gady.book.....B} because we plot the density of states against the relative energy, while \citeauthor{2008gady.book.....B} plot it against energy. The density of states plays a key role in classical statistical mechanics, and also in galactic dynamics, where it allows to compute the {\it differential energy distribution}, $N(E)=g(E)f(E)$, defined in such a way that $N(E)dE$ is the number of stars with energies in the range $E+dE$. Note that the integrand appearing in our Eq.~(\ref{eq:mass_contri}) is basically the differential energy distribution expressed in terms of the relative energy $\epsilon$.  In line with the connection between $\varpi(\epsilon)$ and the density of states, the quantity defined
in Eq.(\ref{eq:master_mind}) would be related to the volume in phase-space per unit energy and per unit radius $r$. In fact, $4 \pi r^2 \xi(\epsilon,r) d\epsilon dr$ is the volume in phase space corresponding to particles with energies in the range $(\epsilon,\epsilon+d\epsilon)$ and radii in the range $(r,r+dr)$. For our purpose, in order to implement the procedure for inferring the distribution function $f(\epsilon)$, we find it convenient to give the quantities $\xi(\epsilon,r)$ and $\varpi(\epsilon)$ an alternative interpretation, as we have already explained. Our interpretation, although different form the classical one, is consistent from a formal point of view and more useful in practice for our purpose.

\section{Testing the numerical calculation of the eigendensities (proper densities)}\label{app:tests}

\citet[][Eq.~A19]{2023ApJ...954..153S} showed that the DF corresponding to a self-gravitating Schuster-Plummer density profile is analytic,
\begin{equation}
 f(\epsilon) = \frac{120}{(2\pi)^{3/2} \Gamma(9/2)} \frac{\rho_{sp}}{\epsilon_{max}^{3/2}}\,(\epsilon/\epsilon_{max})^{7/2},
  \label{eq:theory}
\end{equation}
with $\epsilon_{max}$ and $\rho_{sp}$ defined in Sect.~\ref{sec:eigen_sp} and $\Gamma(x)$ the gamma function.
Formally, it is like the polynomial expansion we use for $f(\epsilon)$ (Eq.~[\ref{eq:polydef}]) with a single term,
\begin{equation}
f(\epsilon)= \frac{a_{\frac{7}{2}}}{\epsilon_{max}^{3/2}}\,(\epsilon/\epsilon_{max})^{7/2},
\end{equation}
where
\begin{equation}
  a_{\frac{7}{2}}=\rho_{sp}\frac{120}{(2\pi)^{3/2} \Gamma(9/2)}.
\end{equation}
On the other hand, the surface density corresponding to a Schuster-Plummer density profile is also analytic
\citep[e.g.,][]{2008gady.book.....B,2022Univ....8..214S},
\begin{equation}
\Sigma(R) = \rho_{sp}r_{sp}\frac{4/3}{[1+(R/r_{sp})^2]^2},
\end{equation}
which must be equal to the surface density provided by Eq.~(\ref{eq:master3}),
\begin{equation}
  \Sigma(R) = a_{\frac{7}{2}} \,S_{\frac{7}{2}}(R),
  \label{eq:app_def}
\end{equation}
which implies 
\begin{equation}
  S_{\frac{7}{2}} (R) = \frac{(2\pi)^{3/2}\,\Gamma(9/2)}{90}\frac{r_{sp}}{[1+(R/r_{sp})^2]^2}.
\label{eq:tests}
\end{equation}

We have used the expression~(\ref{eq:tests}) to test the numerical algorithms developed to compute $S_i(R)$. The result is shown in Fig.~\ref{fig:tests}, which compares the analytic and numerical expressions. The relative difference between them at each radius  ($|\Delta S_i/S_i|$; the red dashed line) is always smaller than 1\,\%, and it is typically smaller than 0.1\,\% in the core ($R/r_{sp} < 1$). In addition, the difference relative to the maximum of the profile is always smaller than 0.1\,\%  ($|\Delta S_i|/\max S_i$; the red dotted line).
\begin{figure}
\centering
\includegraphics[width=\linewidth]{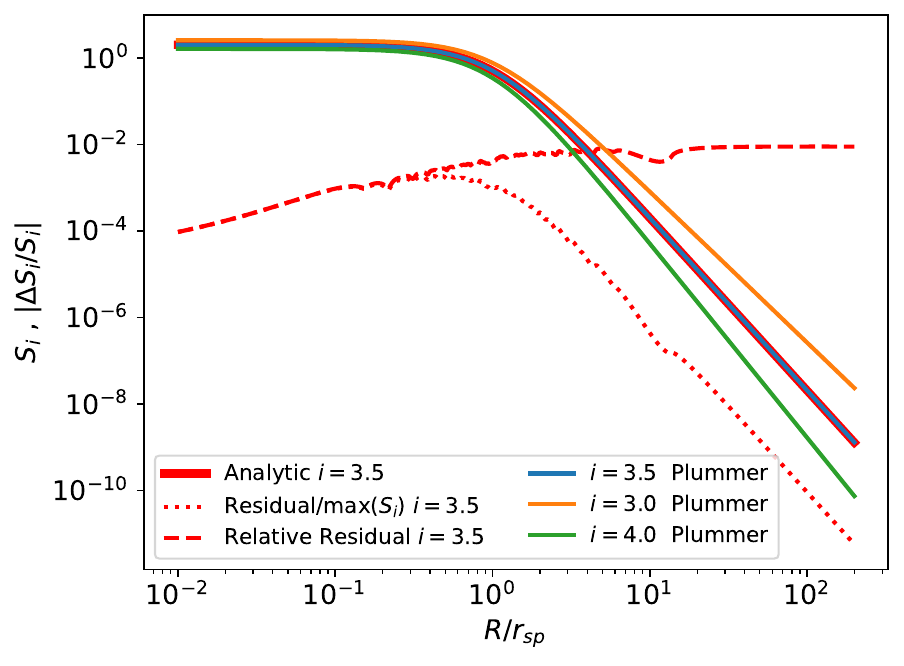}
\caption{Tests for the numerical calculations employed to compute the characteristic densities $S_i(R)$ in Eq.~(\ref{eq:master4}). The case of a self-gravitating Schuster-Plummer profile provides the analytic expression for  $i=7/2$ given in  Eq.~(\ref{eq:tests}). The analytic and numerical expressions are shown as indicated in the inset (marked as $i=3.5$). The differences between them ($\Delta S_i/S_i$ and $\Delta S_i/\max S_i$) are also included as the red dotted and dashed lines. The cases $i=3$ and $i=4$ are shown for reference.}
  \label{fig:tests}
\end{figure}

%
%
\section{Sanity checks to test the algorithm that retrieves DFs}\label{sec:sanity}

This Appendix collects a number of sanity checks that support the robustness and  consistency of the diagnostic method used in the paper (Sect.~\ref{sec:the_actual_algorithm}).

\begin{figure}
\centering
\includegraphics[width=\linewidth]{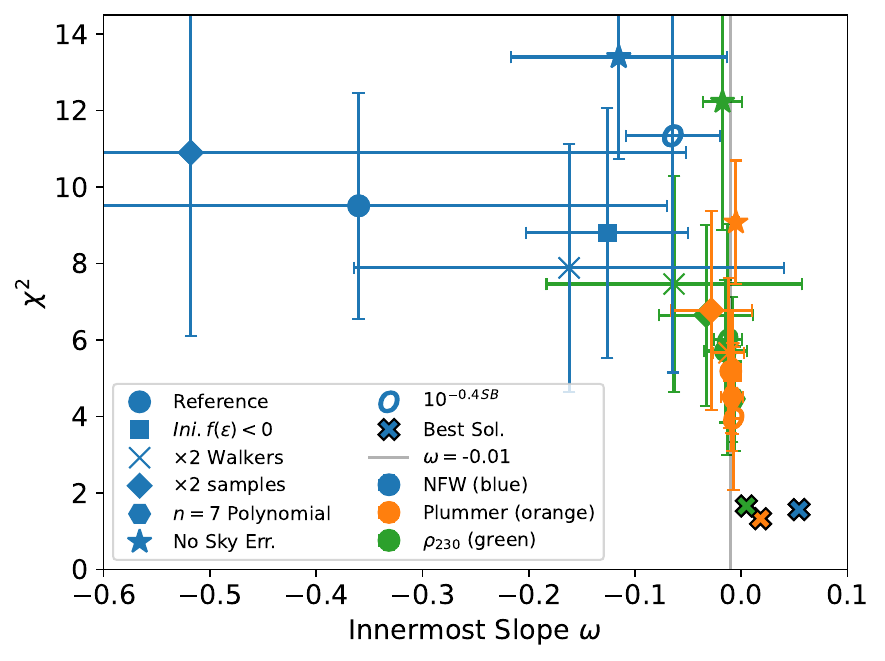}
\caption{
Scatter plot summarizing how changing the hyper-parameters of the fit affect the interpretation of \nube 's  surface density profile. The symbols with error bars represent the mean and the standard deviation of $\chi^2$ and $\omega$ inferred from the different posteriors. Each color corresponds to a different potential, with blue, orange, and green symbols representing NFW, Schuster-Plummer, and $\rho_{230}$ potentials, respectively. Each type of symbol corresponds to a different set of hyper-parameters as indicated in the inset (see Appendixes~\ref{sec:hyper_param} and \ref{app:referee} for details). The values for the nominal hyper-parameters used in the main text are portrayed as bullet symbols and denoted as ``Reference'' in the inset.  The figure also includes the $\chi^2$ and $\omega$ from the best fits obtained with unconstrained $f(\epsilon)$, which are the same as the thick times symbols shown in Fig.~\ref{fig:df4_run_plotf}. 
  The vertical grey line is the same as that in Fig.~\ref{fig:df4_run_plotf}.
}
\label{fig:nube4}
\end{figure}
  \subsection{Varying the hyper-parameters that define the algorithm} \label{sec:hyper_param}
  In order to test the dependence of the results on the hyper-parameters defining the algorithm, \nube 's profile was re-fitted changing them. Among others, these hyper-parameters define the priors of the Bayesian fits. Since the main result stemming from the application of the algorithm is the fact that the gravitational potential of \nube\ arises from a mass distribution with a core rather than a cusp  (Sects.~\ref{sec:potential_nube} and \ref{sec:conclusions}), we study whether this result is modified by the use of hyper-parameters different from the nominal ones in Sect.~\ref{sec:the_actual_algorithm}.  The relationship between $\chi^2$ and the innermost slope $\omega$ is used as diagnostics tool. These two quantities are used in the main text to argue that cuspy NFW potentials provide worst fits than the cored  Plummer-Schuster or $\rho_{230}$ potentials.  Simply put,  $\chi^2$ is larger for cuspy potentials that also provide $\omega$ farther away from the observed value (Sect.~\ref{sec:application} and Fig.~\ref{fig:df4_run_plotf}). Figure~\ref{fig:nube4} shows $\chi^2$ versus $\omega$ for different hyper-parameters. The symbols with error bars represent the mean and the standard deviation of the corresponding distribution of $\chi^2$ and $\omega$ inferred from the posterior. Each color corresponds to a different potential, with blue, orange, and green symbols representing NFW, Schuster-Plummer, and $\rho_{230}$ potentials, respectively. Each type of symbol corresponds to a different set of hyper-parameters. The nominal values used in the main text are portrayed as bullet symbols. These values are changed one at a time to produce other alternative hyper-parameters. The square symbols represent initializing the sampling of the posterior with the unconstrained best fitting least squares, which has $f(\epsilon) < 0$. These fits also restrict the amplitudes defining the DF ($a_i$ in Eq.~[\ref{eq:polydef}]) to within $10^{-2}$ and $10^{2}$ of the best fit values. 
The times symbols correspond to doubling the number of wakers when sampling the posterior.
The diamond symbols correspond to doubling the number of samples when sampling the posterior.
The hexagon represents fits where the order of the polynomial used for $f(\epsilon)$ ($n$ in Eq.~[\ref{eq:polydef}]) differs from the nominal value (7 rather than 10). Other orders of the polynomial give similar results. 
For reference,  Fig.~\ref{fig:nube4}  includes the $\chi^2$ and $\omega$ from the best fits obtained with unconstrained $f(\epsilon)$ which are the same as the thick times symbols shown in Fig.~\ref{fig:df4_run_plotf}. 

The main conclusion arising from inspecting Fig.~\ref{fig:nube4} is that NFW potential fits (in blue) are always worst than  Schuster-Plummer potential fits (in orange) and $\rho_{230}$ fits (in green).  Specifically, the NFW fits exhibit larger  $\chi^2$ and  $\omega$ values  deviating farther from those of \nube, represented in Fig.~\ref{fig:nube4} by the solid time symbols. This systematic preference for cored over cuspy fits is independent of the chosen set of hyper-parameters, reinforcing the robustness of the results against specific details of their selection.

\subsection{Impact of the uncertainties in the mass profile of \nube }\label{app:referee}
\begin{figure}
\centering
\includegraphics[width=\linewidth]{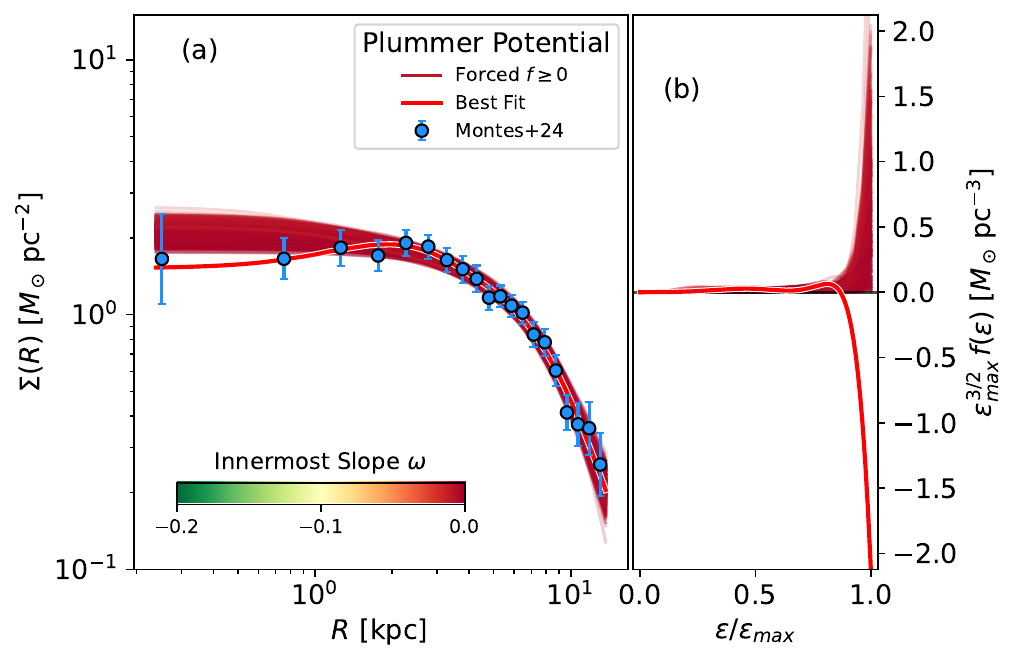}
\includegraphics[width=\linewidth]{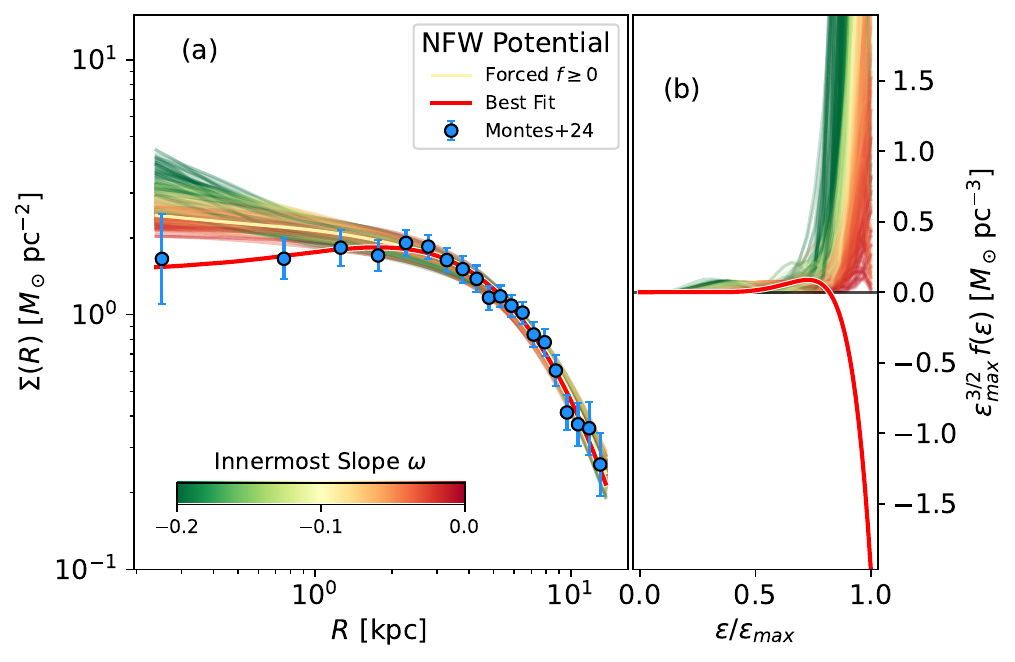}
\caption{Similar to the fits corresponding to Figs.~\ref{fig:df4_run_plot_2} and \ref{fig:df4_run_plot_1}  but with smaller error bars that consider only photometric errors and errors in the mass-to-light ratio calibration. The fits assuming a Schuster-Plummer potential are shown in the top panels whereas those assuming a NFW potential are in the bottom panels. The result discussed in the main text remains.  Compared with the NFW fits, the Schuster-Plummer potential fits with $f(\epsilon)>0$ (colored lines) have both an inner slope closer to the  observed one and a smaller $\chi^2$. The color code is the same in the two panels and also the same of the code used in the figures of the main text. For further details, see the caption of Fig.~\ref{fig:df4_run_plot_2}. 
}
  \label{fig:df41_run_plot_41}
\end{figure}
Here we study two aspects of the uncertainties in  the mass profile of \nube~(Fig.~\ref{fig:nube}) that may potentially have an impact on the conclusions. The first one has to do with the error bars employed in the calculation of $\chi^2$  (Eq.~[\ref{eq:chi2def}]) which affect the posterior and may potentially influence the conclusions. The ones used in the main text and shown in Fig.~\ref{fig:nube} are larger than the scatter of observed points because, together with the Poisson errors associated with the photometry, they include the systematic errors associated to the sky subtraction and the estimate of mass-to-light ratio \citep{2024A&A...681A..15M}. What happens if these other errors are disregarded leaving error bars closer to the scatter of the individual points in the radial profile? We carry out this exercise and the result is shown in Fig.~\ref{fig:df41_run_plot_41}.  Obviously, the overall value of the $\chi^2$ increases, but the conclusions reached in the main text remain.  Compared with the NFW fits,  Schuster-Plummer potential fits having $f(\epsilon)>0$ (the colored lines in Fig.~\ref{fig:df41_run_plot_41}) have an inner slope closer to the observed one together with a significantly smaller $\chi^2$. This fact can be better appreciated in the diagnostic plot shown in Fig.~\ref{fig:nube4}, where the corresponding $\chi^2$ and $\omega$ are shown as star symbols labelled with "No Sky Err".

\begin{figure}
\centering
\includegraphics[width=\linewidth]{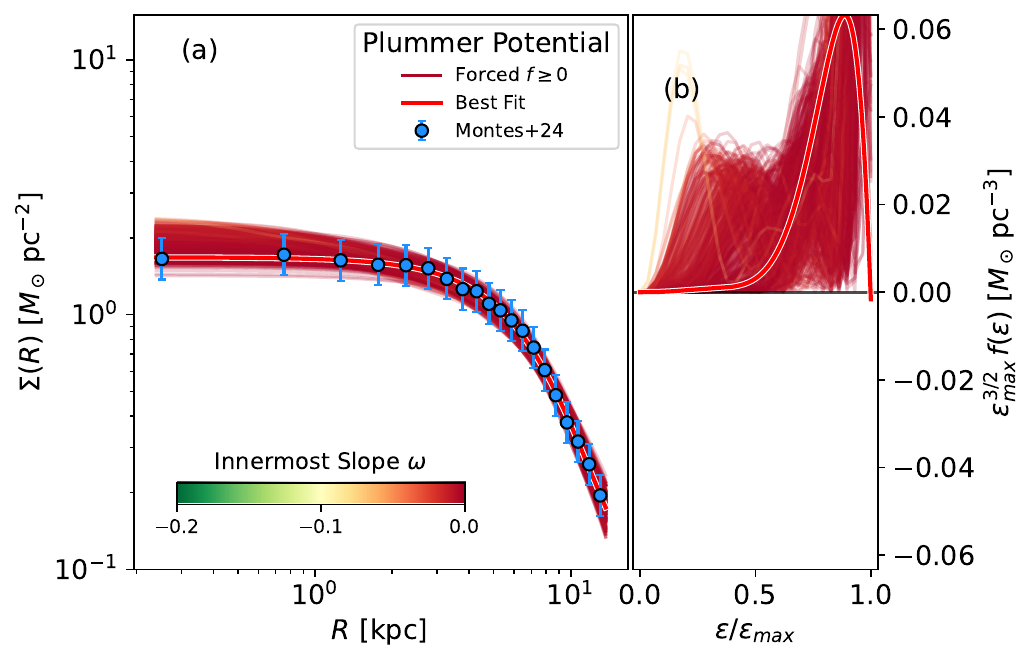}
\includegraphics[width=\linewidth]{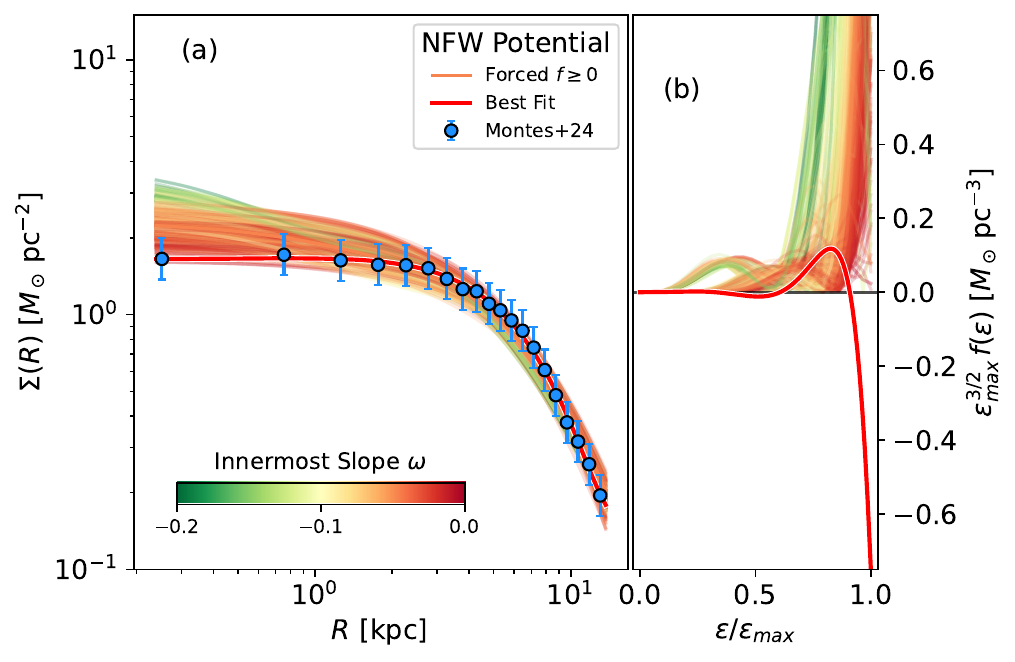}
\caption{
  Equivalent to Figs.~\ref{fig:df4_run_plot_2} and \ref{fig:df4_run_plot_1}  but using a scaled version of the $r$-band surface band profile of \nube\ as mass surface density. This is equivalent to assuming a constant mass-to-light ratio throughout the galaxy.  Note that the central drop in mass has gone away (cf. Fig.~\ref{fig:nube} with the data points in figures {\em a}). The fits assuming a Schuster-Plummer potential are shown in the top panels whereas those assuming a NFW potential are in the bottom panels. This other alternative calibration of the mass profile does not modify the conclusions based on the original calibration. The color code is the same in the two panels and also the same of the code used in the figures of the main text. For further details, see the caption of Fig.~\ref{fig:df4_run_plot_2}. Note that the EIM technique is insensitive to a global scaling of the mass profile, and this arbitrary factor has been chosen here so that the level of $\Sigma(R)$ is similar to the original one in Fig.~\ref{fig:nube}.
}
  \label{fig:df41_run_plot_51}
\end{figure}
The second study refers to the impact of the mass-to-light ratio calibration. As it is mentioned in Sect.~\ref{sec:potential_nube_analytic}, the central drop in mass in Fig.~\ref{fig:nube} is not present in the light profile of \nube\ and may be an artifact appearing when transforming the observed photometry into stellar mass. We consider the impact of the used mass-to-light ratio on the conclusion by using a constant value rather than the color-varying ratio employed by \citet{2024A&A...681A..15M} and used in our study. In this test we use, 
\begin{equation}
\Sigma(R)\propto 10^{-0.4\, {\rm SB}},
\end{equation}
with SB the observed surface brightness. A profile thus compute is shown in Fig.~\ref{fig:df41_run_plot_51}, where we have chosen the $r$-band photometry because the dependence of the mass-to-light ratio on colors is smaller in the red, yet the $r$-band exhibits low noise. The chosen errors are somewhat arbitrary trying to account for photometric errors and matching those in Fig.~\ref{fig:df41_run_plot_51}.  This mass profile do not show the central drop of Fig.~\ref{fig:nube}. Using this data, we repeat the analysis and the resulting fits and DFs are shown in Fig.~\ref{fig:df41_run_plot_51}.  As for the original data set, Schuster-Plummer potential fits are better than the NFW potential fits. They have an inner slope closer to the observed one and their  $\chi^2$ is significantly smaller. Their values are included in Fig.~\ref{fig:nube4} using the symbol $O$ and labelled as $10^{-0.4\,{\rm SB}}$. In addition, the Schuster-Plummer potential best fit does not require the negative distribution function needed for the original \nube\ profile: compare the red thick line in Fig.~\ref{fig:df41_run_plot_51}b (top panel) with  Fig.~\ref{fig:df4_run_plot_2}b. 

Continuing with the impact of changing the mass-to-light ratio, we construct another mock profile using the \nube\ data but re-arranging the order of the observed $\Sigma(R)$ so that the profile monotonically decreases outward in the inner part. Note that such re-arrangements leaves a profile consistent with the original data set keeping in mind the large error bars (Fig.~\ref{fig:nube}). The resulting profile is shown in Fig.~\ref{fig:df4_run_plot_9}. This figure is similar to Fig.~\ref{fig:df2_run_plot_6}, and evidences how the Schuster-Plummer potential does an excellent job with $f(\epsilon)\geq 0$ whereas the profiles forced to have $f(\epsilon) > 0$ in NFW potentials provide much worst fits.  
  \begin{figure}
\centering
\includegraphics[width=\linewidth]{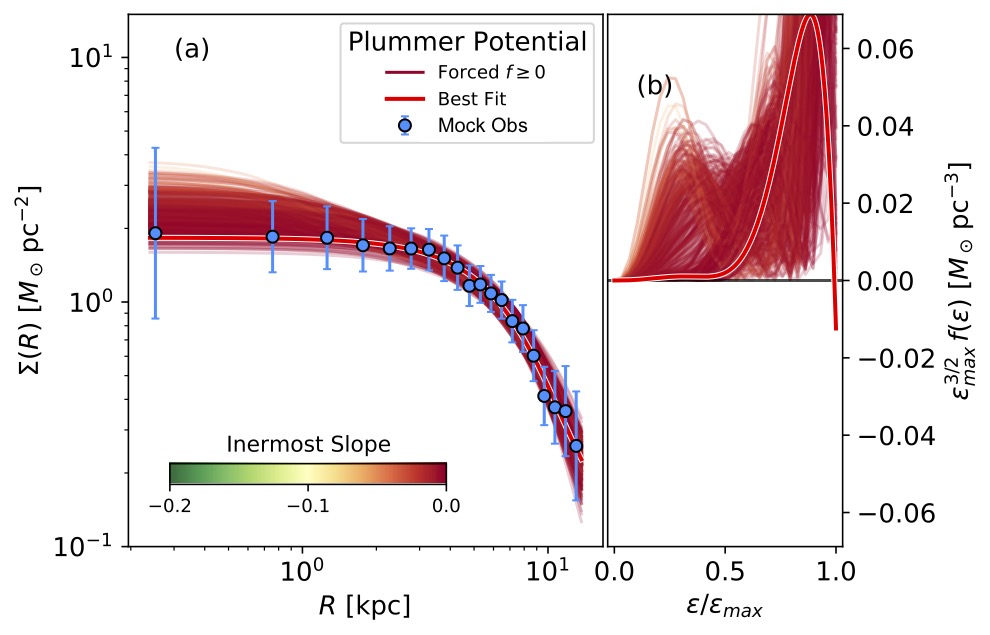}
\includegraphics[width=\linewidth]{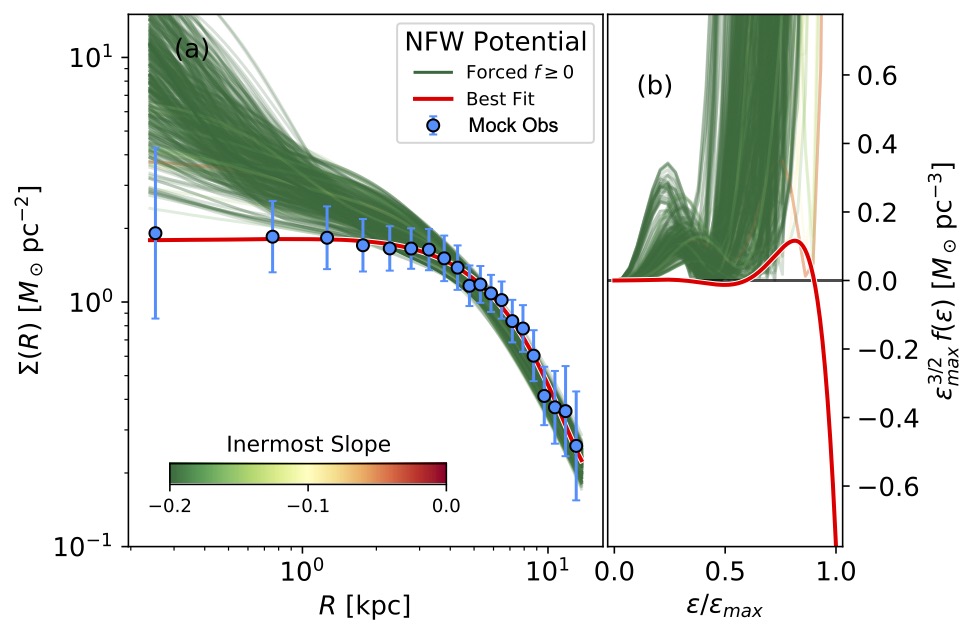}
\caption{
  Mock profiles using all the data from \nube\ but rearranging the inner points so that the resulting profile decreases monotonically outward (the symbols labeled as Mock Obs). This re-arrangement is consistent with the error bar of the observation but clears out the problem of the algorithm we employ to explain positive inner slopes.  Top panels: case of a  Schuster-Plummer potential. Bottom panels:  case of a  NFW potential. This figure is similar to Figs.~\ref{fig:df2_run_plot_6}, having identical color code.}
\label{fig:df4_run_plot_9}
\end{figure}
%

%

\medskip
\subsection{Recovering properties of known core profiles}\label{sec:recovery}
\begin{figure}
\centering
\includegraphics[width=\linewidth]{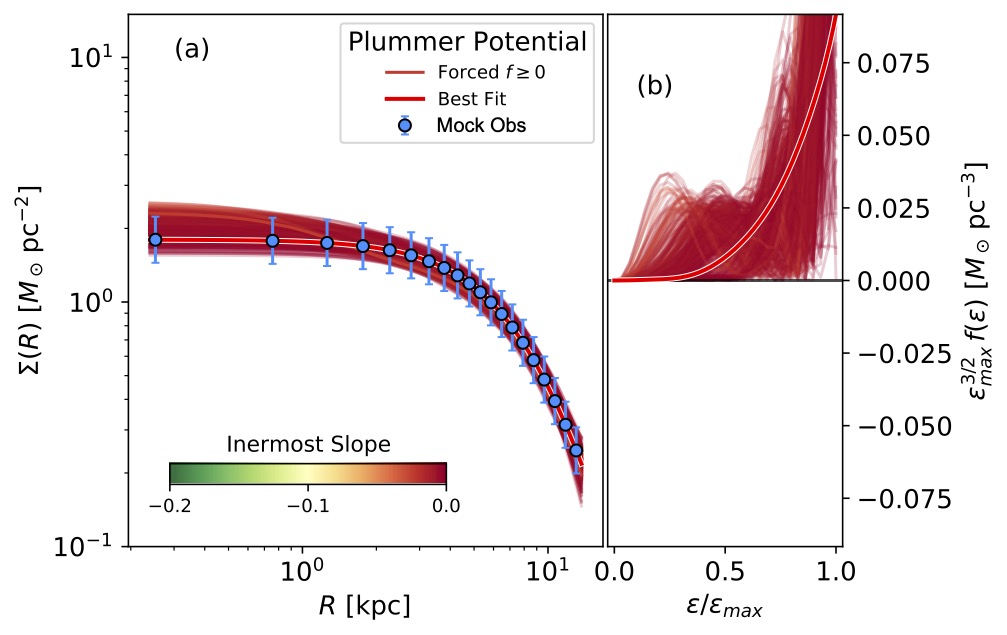}
\includegraphics[width=\linewidth]{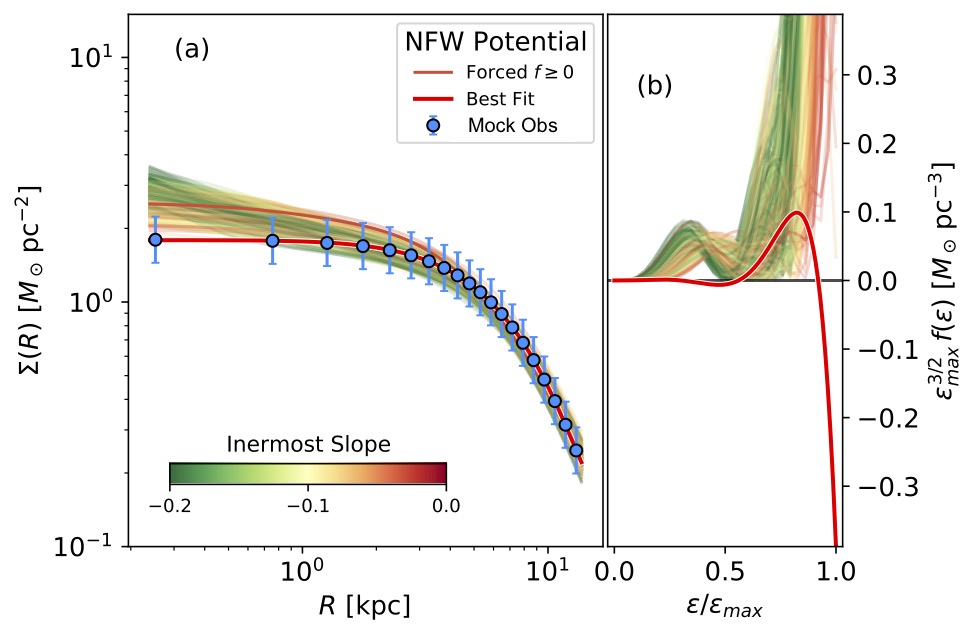}
\caption{Summary of the analysis carried out with a mock mass surface density profile corresponding to a Schuster-Plummer profile (the symbols labeled as Mock Obs). The figures are similar to Figs.~\ref{fig:df4_run_plot_2}, \ref{fig:df4_run_plot_1}, and \ref{fig:df4_run_plot_3} in the main text. The result assuming a Schuster-Plummer potential is shown in the top panels whereas the result assuming a NFW potential is in the bottom panels. Note how the free-$f(\epsilon)$ fits (the thick red lines) yield $f(\epsilon)>0$ for the Schuster-Plummer potential whereas it requires $f(\epsilon)<0$ for the NFW potential. The fits forced to have $f(\epsilon)>0$ (color thin lines) are much worst in the case of the NFW profile. The color code is the same in the two panels and also the same of the code used in the figures of the main text.
}
  \label{fig:df2_run_plot_6}
\end{figure}
We repeat the analysis with a mock mass surface density profile corresponding to a Schuster-Plummer profile with  error similar to those of \nube . The result assuming a Schuster-Plummer potential is shown in Fig.~\ref{fig:df2_run_plot_6}, top panels. Obviously, a Schuster-Plummer profile is fully consistent with a Schuster-Plummer potential and, as expected, the inferred $f(\epsilon)$ (the red solid line) is always positive and in agreement with its expected form \citep[][given explicitly in Eq.~\ref{eq:theory} above]{2023ApJ...954..153S}. Moreover, the surface density profiles and DFs resulting from the MCMC exploration of the posterior forcing $f(\epsilon)\geq 0$ are also compatible with this best fit. This test shows what is to be expected in case of assuming a potential fully consistent with the observed $\Sigma(R)$. On the other hand, the same mock surface density is analyzed assuming a NFW potential (Fig.~\ref{fig:df2_run_plot_6}, bottom panels). The required $f(\epsilon)$ is negative, and so unphysical, and the fits forcing $f\geq 0$ are way off the best fitting profile.

\end{document}